\documentclass[usenatbib]{mnras}

\usepackage{multirow}
\usepackage{longtable}
\usepackage[T1]{fontenc}
 \usepackage{ae,aecompl}
\usepackage{amsmath}
\usepackage{breqn}
\usepackage{graphicx}
\usepackage{textcomp}
\usepackage{subfigure}
\usepackage{amssymb}    
\PassOptionsToPackage{authoryear}{natbib}
\usepackage{natbib}
\bibpunct{(}{)}{;}{a}{}{,}
\PassOptionsToPackage{english}{babel}
\usepackage{babel}

\PassOptionsToPackage{hyphens}{url} 
\usepackage{amsbsy}
\usepackage{amssymb}
\usepackage{amsmath}
\usepackage{graphicx}
\usepackage{color}
\usepackage{xcolor}
\usepackage{natbib}

\newcommand{\be}{\begin{equation}}
\newcommand{\ee}{\end{equation}}
\newcommand{\bear}{\begin{eqnarray}}
\newcommand{\eear}{\end{eqnarray}}

\def \aap {AAP}
\def \apj {ApJ}
\def \apjl {ApJL}

\def \mnras {MNRAS}
\def \prd {Phys. Rev. D}

\begin{document}

\title[]{A novel approach for the analysis of the geometry involved in determining light curves of pulsars}
\author[Vigan\`o \& Torres]{Daniele Vigan\`o$^{1,2,3,4}$, \& Diego F. Torres$^{4,5,3}$ \\
$^1$Departament  de  F\'{\i}sica, Universitat  de  les  Illes  Balears, Palma  de  Mallorca,  Baleares  E-07122,  Spain\\
$^2$Institut Aplicacions Computationals (IAC3),  Universitat  de  les  Illes  Balears,  Palma  de  Mallorca,  Baleares  E-07122,  Spain\\
$^3$Institut d'Estudis Espacials de Catalunya (IEEC), 08034 Barcelona, Spain\\
$^4$Institute of Space Sciences (ICE, CSIC), Campus UAB, Carrer de Magrans s/n, 08193 Barcelona, Spain\\
$^5$Instituci\'o Catalana de Recerca i Estudis Avan\c{c}ats (ICREA), Barcelona, Spain
}

\date{}

\pagerange{\pageref{firstpage}--\pageref{lastpage}} 
\pubyear{2018}

\maketitle

\label{firstpage}

\begin{abstract}
In this work, we introduce the use of the differential geometry Frenet-Serret equations to describe a magnetic line in a pulsar magnetosphere. 
These equations, which need to be solved numerically, fix the magnetic line in terms of their tangent, normal, and binormal vectors at each position, given assumptions on the radius of curvature and torsion. 
Once the representation of the magnetic line is defined, we provide the relevant set of transformations between reference frames; the ultimate aim is to
express the map of the emission directions in the star co-rotating frame. 
In this frame, an emission map can be directly read as a light curve seen by observers located at a certain fixed angle with respect to the rotational axis. 
We provide a detailed step-by-step numerical recipe to obtain the emission map for a given emission process, and give a set of simplified benchmark tests.
Key to our approach is that it offers a setting to achieve an effective description of the system's geometry {\it together} with the radiation spectrum. 
This allows to compute multi-frequency light curves produced by a specific radiation process (and not just geometry) in the pulsar magnetosphere, and intimately relates with averaged observables such as the spectral energy distribution.
\end{abstract}


\section{Introduction}

About 250 neutron stars show detectable periodicity in the $\gamma$-ray range, with spinning periods from milliseconds to seconds. 
Detections include all pulsar classes except magnetars \citep{2fpc,Abdo2010m,Li2017}.

Ideally, in order to understand the spectral and geometrical properties of the pulsar emission, one has to consider the pulsar rotation, the inclination angle (the angle between the magnetic moment and the rotation), the global electromagnetic field distribution, the spatial distribution of the charged particles, their trajectories, and the spectral and angular distribution of the emitted radiation at each point of their trajectories.

It is not a surprise, then, that due to the complexity of the problem, the modelling of the pulsars' observable properties has usually considered the light curves and the spectral energy distribution separately. 
On the one hand, light curves were usually computed from a purely geometrical perspective, where the acceleration region is assumed to be localized in some specific regions of the magnetosphere, the photon flux emitted by the accelerated particles is assumed to be constant throughout the region, and its dependence with energy is not studied (e.g., \cite{Watters2009,Venter2009,Bai2010b,romani10,pierbattista12,Cao2019}).
In these studies, an exact solution of the force-free, rotating  dipole-dominated magnetosphere is commonly taken as a background on top of which an accelerating region (i.e., a breakdown of the force-free approximation) is placed by hand (for instance, close to the last open field line). 
An alternative has appeared with the FIDO models (force-free inside, dissipative outside) introduced by \cite{Kalapotharakos2014}, see e.g., \cite{brambilla15,Cao2019}.
Only recently,  particle-in-cell simulations tried to identify more consistently the appearance of possible accelerating regions, which appear to be located 
usually beyond the light cylinder (e.g., \cite{Cerutti2016,Philippov2018,kalapotharakos18}, and for a perspective on such recent works and on what have we learned see \cite{Cerutti2018}).
These latest works accurately solve the magnetic configuration for a given set of parameters (inclination angle, light cylinder, magnetic field at the surface). However, due to the computational costs and the ad-hoc assumptions made, they are neither easily usable for fitting the light curves of a specific pulsar, nor to address the variety of the known sources. 

On the other hand, the spectral distributions have been described by synchro-curvature radiation (or simply curvature), see, e.g., \cite{romani96,zhang97,hirotani99a,takata08,takata17,hirotani15,paper4,Petri2019}, or inverse Compton, see e.g., \cite{hirotani99b,lyutikov12}. 
In our previous works, we have used the full synchro-curvature radiation formulae 
(providing a more compact formulation \citep{paper0} than the one originally introduced by \cite{cheng96}).
After general considerations about the usually overlooked strong assumptions of the so-called outer gap \citep{paper1}, we have presented a new emission model \citep{paper2} based on a few effective parameters that allow to calculate the trajectories of the particles in a generic accelerating region, together with the related emitted radiation. 
In this way, we have successfully fitted the $\gamma$-ray radiation of the observed pulsars \citep{paper3,paper4}. This effective, reverse-engineering approach allows to infer physical quantities defining the spectrum, like the accelerating electric field or the magnetic gradient, instead of fixing them a-priori, and then to find correlations among them.
A recent extension of the model to encompass also the X-ray regime \citep{torres18} has allowed to fit the spectra of pulsars across an energy range of about ten orders of magnitude with just three parameters. 
We already showed how these  spectral models are predictive, since from the $\gamma$-ray best fit alone, one can infer the plausible luminosity in X-rays in most of the cases. 
Such $\gamma$-ray-driven search of previously unseen X-ray pulsators has already led to new detections \citep{li18}.
Additionally, our model has also been recently applied to the full sample of non-thermal X-ray/$\gamma$-ray pulsars, a sizeable sample 
formed by 40 members \citep{CotiZelati2019}, finding a good agreement in all cases, with minimal conceptual extensions requirements \citep{Torres2019}.

However, in all our earlier works, we have ignored the geometry and the angular distribution of the radiation emitted.
In this paper, then, we lay the foundations for performing pulsar light curves calculations coupled to our spectral emission models. 
We intend to proceed with the same idea: we want to isolate the minimum set of assumptions so that a meaningful light curve prediction can be made. 
The ultimate goal will be to obtain a light curve {\it concurrently} with a spectral prediction, in a way that is versatile enough so that we can apply it to describe the many well-characterized pulsars we know (or shall know with future surveys), 
as well as to make multi-frequency prognosis even when based on partial sets of data. 
%
%
We aim to an approach in which one considers a pulsar with known timing properties, and sets a few input parameters: the curvature of the field lines, the value of the magnetic field along them, and the accelerating field. Starting from that, one can calculate the emission map of the sky, considering point by point the direction and spectra of the emitted radiation.

\section{Geometric model}

\subsection{Differential geometry of the lines}

In order to have an effective description of the system's
geometry together with the radiation spectrum, we shall employ the Frenet--Serret  differential geometry formulae to describe particles' trajectories. 
The Frenet--Serret equations are introduced here to describe the geometry of a line given the radius of curvature and torsion as functions of the position.
The derivatives of the tangent $\hat{t}$, normal $\hat{n}$ (pointing towards the curvature center), and the binormal ($\hat b= \hat{t}\times\hat{n}$) unit vectors are expressed in these formulae in terms of each other.
First, we consider a line parametrized by $\lambda = l/R_{lc}$,\footnote{We explicitly note here that we have called $x$ the coordinate along the trajectory in our previous papers, but we allow us to change the name convention here from $x$ to $\lambda$ in order to avoid confusion with the coordinate grid used later in this paper.} i.e. the position of the particle measured along the field line, normalized by the light cylinder $R_{lc} = cP/2\pi = c/\Omega$, where $\Omega$ is the angular spin velocity of the star considered, and $P$ its spin period. 
Second, we need the curvature radius $r_c(\lambda)$ and the torsion $\tau(\lambda)$ along the line, as functions of the position, in order to solve the following set of equations:
\begin{eqnarray}
&& \frac{d\hat{t}}{d\lambda} = \frac{1}{r_c}~\hat{n} \label{eq:fs1}~ , \\
&& \frac{d\hat{n}}{d\lambda} = -\frac{1}{r_c}~\hat{t} + \tau~\hat{b} \label{eq:fs2}~ , \\
&& \frac{d\hat{b}}{d\lambda} = -\tau~\hat{n} ~.
\end{eqnarray}
We shall use a fourth-order Runge-Kutta method to properly integrate the curves. 

For simplicity, and in order to avoid introducing a further parameter, we shall assume that the torsion $\tau$ is zero, which is an acceptable approximation if the twist of the line is negligible compared to its curvature. This allows us to consider that the lines are contained in a 2D plane.
In case one wants to consider the toroidal field $B_t$, one needs to include the torsion as well, and consider the three non-zero components of the direction. We leave this out of our treatment for the moment, but can be easily included, if concurrently considering the additional degrees of freedom.
The zero torsion assumption might end up being not realistic in a twisted magnetosphere. Close to the light cylinder, and for some lines at least, the overall twist may become important, as numerical simulations show, see e.g.,  \cite{spitkovsky06}. On the other hand, though, from what we know after our spectral-only model (see \cite{torres18,Torres2019}), the relevant region of emission is always small in comparison with the scale of the light cylinder. In these small regions, the torsion of the lines can indeed be negligible, even if it is not negligible for the whole line.
But in fact, we simply do not know whether torsion is essential for reproducing the real pulsar light curves observed. 
Thus, it seems appropriate to try a simpler model first and only add the torsion complexity if needed, later.

\subsection{Emission distribution in the local particle reference frame (PRF)}\label{sec:prf}

For a given point $\lambda$ along the trajectory, we can consider a local Particle Reference Frame (PRF), such that the local values of the tangent and normal of the line, $\hat{t}(\lambda)$ and $\hat{n}(\lambda)$, are given by $(\theta_p,\phi_p)=(0,0)$ and $(\pi/2,0)$, respectively, or $(x_p,y_p,z_p) = (0,0,1)$ and $(1,0,0)$ in Cartesian coordinates.

Let us consider the radiation locally emitted by a particle.
When particles travel with a Lorentz factor $\Gamma$, the angular distribution of their emission is spread, but it sharply peaks at the boundary of a cone, centered around the direction of motion. The opening of such cone is energy-dependent, with less energetic photons being more spread than the more energetic ones. However, such opening angle is of the order of $\sim 1/\Gamma$, which, in our scenario, is always a negligible value ($\lesssim 10^{-4}$), compared with the other angular size values at play.

Therefore, we shall consider that all the photons are emitted in the instantaneous direction of motion of the particle, regardless of their energy. Parameterizing the gyration angle with $\chi \in [0,2\pi]$, the instantaneous emission direction in the PRF is therefore given by
\begin{eqnarray}
&& \theta_e(\lambda) = \alpha(\lambda)~,\label{eq:angles_cone}\\
&& \phi_e(\chi) = \chi~,\label{eq:angles_cone_phi}
\end{eqnarray}
where $\alpha(\lambda)$ is the particle pitch angle, which is the angle between the spiraling trajectory around the magnetic field line and the line tangential direction $\hat{t}$. Since the gyration period is much smaller than any relevant time scale (i.e., the region crossing time, the instrumental resolution...), and many particles are supposedly emitting at the same time, we shall consider the emission distribution integrated over a gyration period. Therefore the gyro-averaged emission distribution, for a given $\lambda$, describes a circle centered around $\hat{t}(\lambda)$ in the PRF unit sphere, with a radius given by the particle pitch angle.

In order to have an expression for $\alpha(\lambda)$, one need to consider the equations of motion for the particle relativistic momentum $\vec{p}$, related to the accelerating electric field $E_\parallel$ and to the synchro-curvature power, $P_{sc}$, by: 
\begin{equation}
\label{motion}
{d\vec{p}}/{dt} = eE_\parallel \hat{b} - ({P_{sc}}/{v})~\hat{p},
\end{equation} 
where $e$ is the particle charge and $v$ its velocity modulus. In \cite{paper0} and following works, we evolved the parallel and perpendicular momenta of particles, considering how $P_{sc}$ depends on the Lorentz factor $\Gamma$, $\alpha$ itself, the local value of the magnetic field $B$, and $r_c$. For a given parameterization of $E_\parallel$, $B$, $r_c$, one can consistently evolve the pitch angle and Lorentz factor values along the trajectory calculation (see examples and details in \citet{paper0}).
In general, the pitch angle can have a sizeable value at the beginning of its acceleration phase. In this first stage, the emission is close to the standard synchrotron emission. However, the pitch angle soon tends (exponentially) to zero due to the perpendicular moment losses, so that the whole radiation will be essentially directed along the field line (tangent direction $\hat{t}$, $\theta_e=0$), and the emission can be 
approximated well by the standard curvature formulae.

\begin{figure*}
\centering
\includegraphics[width=.8\textwidth]{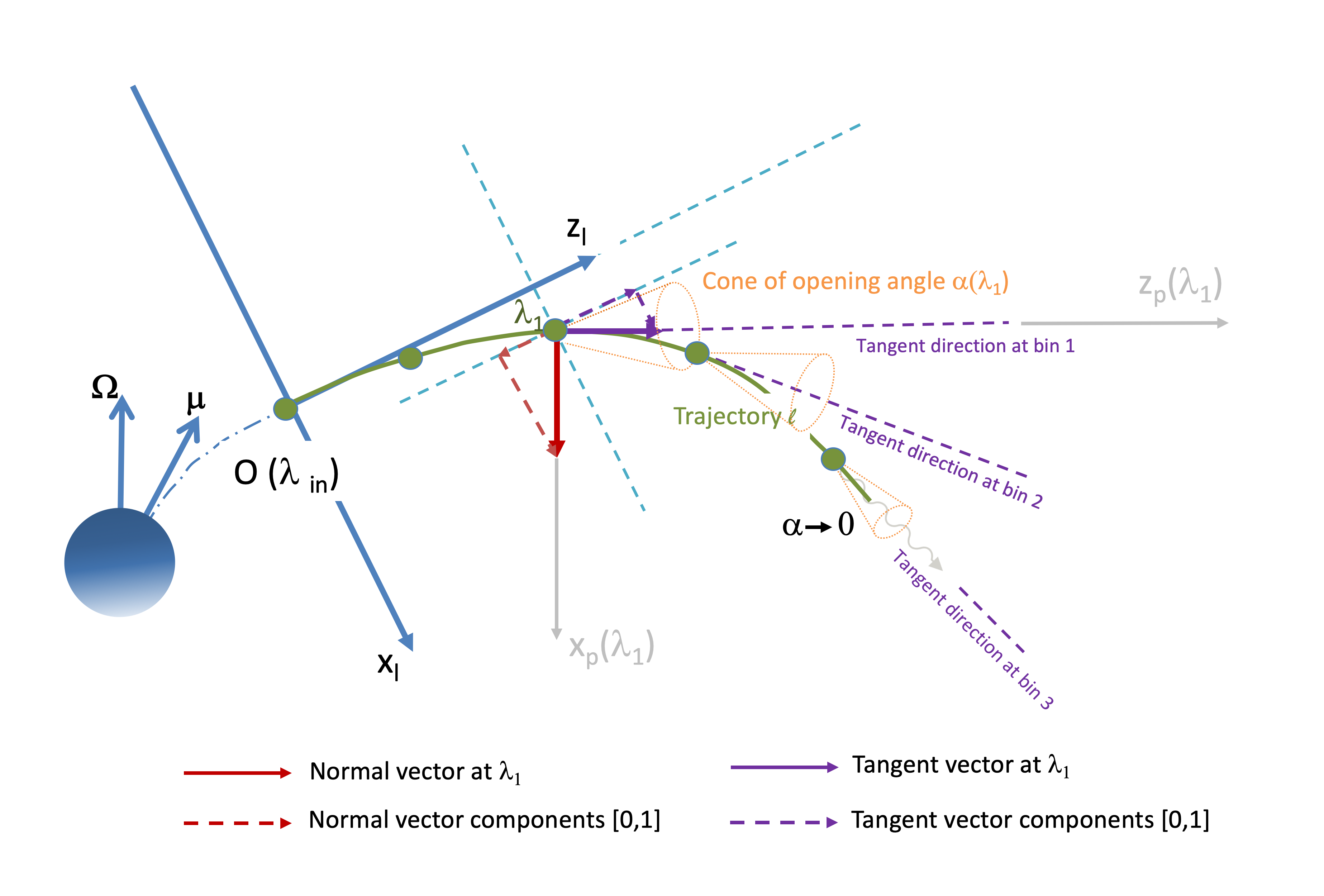}
\caption{Geometrical definitions related to the particle (PRF) and line (LRF) reference frames. See text for a discussion.  }
\label{geo}
\end{figure*}

Fig.~\ref{geo} helps describe the geometry considered and is further relevant for the next section, where we consider how to convert the PRF emission directions into other reference frames.

\subsection{Change of coordinates}

Ultimately, we want to express the map of the emission directions in the star co-rotating frame. 
Starting from the latter, and for a given radiative process, an emission map can be used to generate  
a light curve as seen by observers located at a fixed angle with respect to the rotational axis.
%
For this we have to take into account:
\begin{itemize}
	\item the two angles of the inner point of the accelerating region with respect the magnetic moment $\vec{\mu}$ (magnetic co-latitude and longitude);
	\item the angle between the magnetic moment $\vec{\mu}$ and the rotational axis (inclination angle);
	\item the time-of-flight delay and relativistic aberration effects.
\end{itemize}
Note that we ignore any involved translations that may be needed in changing reference frames. This is because the distances between the frame origins (the size of the magnetosphere at most, i.e. typically up to thousands of km) are negligible compared to the distance from the source to the observers (typically several kpc). Observers
are effectively taken to be 
at infinity.  Below, we explicitly describe the rotations one by one.

%

\subsubsection{Rotation matrices}

If we consider the unit sphere (or sky map) described by two angles, the co-latitude $\theta \in [0,\pi]$ and the longitude $\phi \in [0,2\pi]$, the Cartesian coordinates as a function of the co-latitude and longitude are given by:
\begin{eqnarray}
&& x = \sin\theta\cos\phi~, \nonumber \\
&& y = \sin\theta\sin\phi~, \label{eq:unit_sphere} \\
&& z = \cos\theta~, \nonumber 
\end{eqnarray}
while the inverse relations are
%
\begin{eqnarray}
&& \theta=\arccos (z) , \label{eq:theta_def} \\
&& \phi = {\rm sign}(y)\arccos \left(\frac{x}{(x^2+y^2)^{1/2}}\right)~. \label{eq:phi_def}
\end{eqnarray}
In order to perform all the necessary changes of coordinates, we shall first quite generally consider a unit vector, for which the coordinates $\hat{r}_a$ in the frame A are generally described by a frame B by means of the application of a rotation on the vector $\hat{r}_a$ (hereafter we indicate with a subscript the frame to which the coordinates refer to). A rotation around the $z$-axis by an angle $\xi$ (azimuthal rotation) is defined by a matrix
\begin{equation}\label{eq:rotation_z}
{\bf R}^z(\xi) =
\begin{bmatrix}
\cos\xi & \sin\xi & 0 \\
- \sin\xi & \cos\xi & 0 \\
0 & 0 & 1 
\end{bmatrix} ,
\end{equation}
and a rotation around the $y$-axis by an angle $\Psi$ (meridional rotation) is defined by a matrix
\begin{equation}\label{eq:rotation_y}
{\bf R}^y(\Psi) =
\begin{bmatrix}
\cos\Psi  & 0 & -\sin\Psi \\
0 & 1 & 0 \\
\sin\Psi & 0 & \cos\Psi
\end{bmatrix} .
\end{equation}
A general rotation can then be described as a combination of two matrices, remembering that:\\

(i) their inverse are simply ${\bf R}^z(-\xi)$ and  ${\bf R}^y(-\Psi)$, thus obtained changing the sign of the off-diagonal matrix elements, \\

(ii) rotations commute only if they are performed around the same axis, i.e., ${\bf R}^z(\xi_1){\bf R}^z(\xi_2)={\bf R}^z(\xi_1 + \xi_2)$ and ${\bf R}^y(\Psi_1){\bf R}^y(\Psi_2)={\bf R}^y(\Psi_1 + \Psi_2)$. \\

In the following, we shall operate three times the same transformations in order to align the $z$-axis to the following reference vectors: 
the tangent to the magnetic line at the beginning of the accelerating region, the magnetic moment $\vec{\mu}$, and the rotational vector $\vec {\Omega}$. 
In each frame, we can describe the unit sphere using the spherical coordinates $\theta$ and $\phi$, useful to visualize maps and for practical purposes, or the Cartesian components,  given by eq.~(\ref{eq:unit_sphere}), over which we can apply directly the rotation matrices (\ref{eq:rotation_z})-(\ref{eq:rotation_y}).

\subsubsection{Line reference frame (LRF)}

The first frame considered is the line reference frame (LRF), also sometimes referred to as locally co-rotating frame, in which the line is still. 
We choose the coordinates so that the tangent line at the innermost point of the region, $\lambda_{\rm in}$, is along the axis $z_l$, and the normal is along $x_l$ (see Fig. \ref{geo}). Therefore, the projection of the tangent to the line in the LRF sky map is:
\begin{eqnarray}
&& \Psi_t(\lambda) = \arccos(z_t(\lambda)) \label{eq:tangent} ~, \\
&& \xi_t(\lambda) = 0 ~, \label{eq:xi_lrf}
\end{eqnarray}
where $x_t$ and $z_t$ are evolved according to the Frenet--Serret formulae and, as mentioned above, we consider that the line has no torsion, so that it is contained in a 2D plane identified by $\phi_l=0$ (or $y_l=0$), which is defined so that it contains the magnetic dipolar moment $\vec{\mu}$.

In order to transform from the PRF to the LRF, one has to perform a rotation around the $y_p$-axis, by $\Psi_t(\lambda)$:
\begin{equation}
\hat{r}_l = {\bf R}^{y}(-\Psi_t) \hat{r}_p~.
\end{equation}
Note, that, in the LRF, the emission cone distribution of the instantaneous radiation is spread around a circle of angular radius $\alpha(\lambda)$, centered at the angle $(\theta_e,\phi_e)=(\Psi_t(\lambda),0)$.

\subsubsection{Magnetic reference frame (MRF)}\label{sec:mrf}

\begin{figure*}
	\centering
	\includegraphics[width=.6\textwidth]{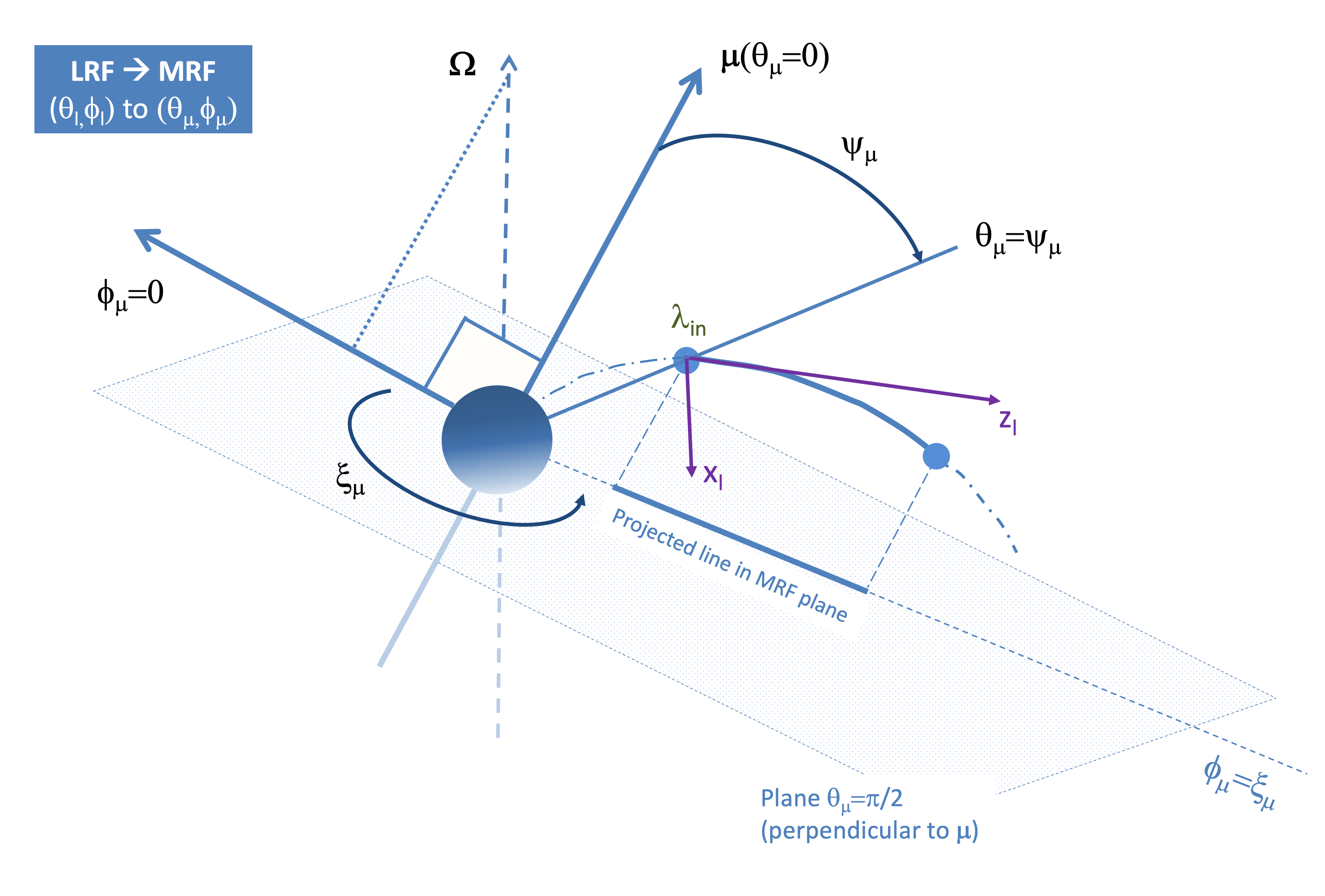}
	\includegraphics[width=.6\textwidth]{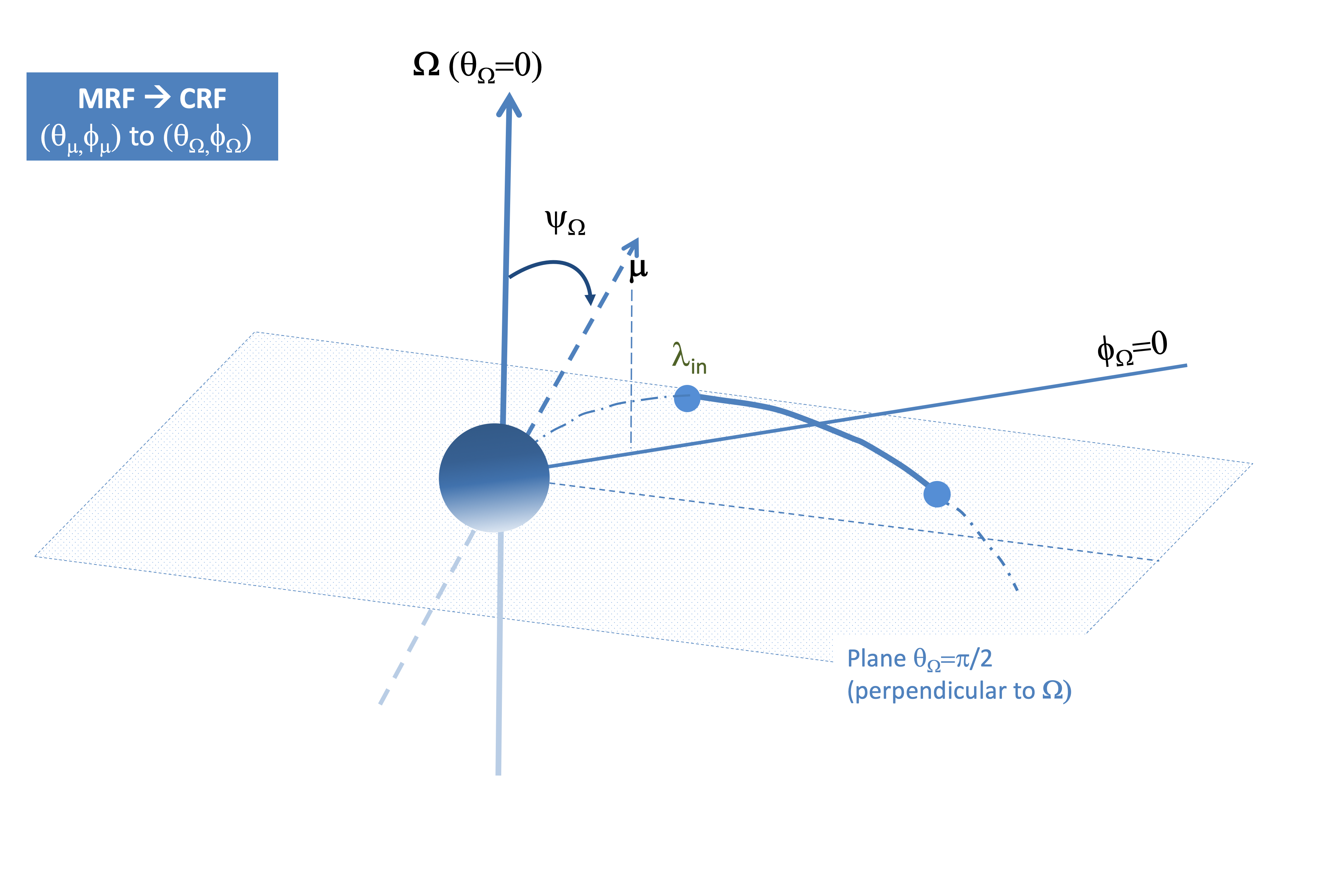}
	\caption{Geometrical definitions related to the transformation from the line to the magnetic reference frame (first panel)
	and from the magnetic to the co-rotating reference frame (second panel). See text for a discussion.}
	\label{geo2}
\end{figure*}

Let us now consider the transformation from the LRF to the magnetic reference frame (MRF), where the magnetic moment is aligned with the $z_\mu$-axis. The top panel of
Fig. \ref{geo2} is a useful reference for what follows.

First, we have to fix the magnetic co-latitude of the inner point of the accelerating region, i.e. the angle $\Psi_\mu$ between its physical position $\vec{x}(\lambda_{in})$ and the magnetic axis of the star $\mu$, measured from the center of the star. 
Second, $\xi_\mu$ is defined as the angle, measured on the plane perpendicular to $\mu$ from $\phi_\mu=0$, i.e. between the projections of the given magnetic field line and $\Omega$.\footnote{Note that, for an aligned rotator, the projection of $\Omega$ becomes singular, and the definition of $\phi_\mu=0$ coincides then with $\phi_\Omega=0$.} Therefore, we need two inverse 
rotations to identify a certain line in the MRF: a meridional one (the magnetic co-latitude $\Psi_\mu$) around $y$ and an azimuthal one (the magnetic longitude $\xi_\mu$) around $z$.
The transformation to pass from the LRF to the MRF is given by
\begin{equation}
 \hat{r}_\mu = {\bf R}^{z}(-\xi_\mu){\bf R}^{y}(-\Psi_\mu) \hat{r}_l ~.
\end{equation}
In these coordinates, the magnetic axis coincides with $z_\mu$, and $y_\mu$ is perpendicular to the plane defined by $\vec{\mu}$ and $\vec{\Omega}$.

If we consider different lines, or a bundle of them (a finite accelerating region), the picture gets formally more complicated. A simplifying model assumption consists in considering that for all lines the trajectory has the same coordinates, Eqs.~(\ref{eq:tangent})-(\ref{eq:xi_lrf}), in their respective LRF. The approximation comes from the idea that the bundle is limited in its transversal direction, so that the range of $\xi_\mu$ and $\Psi_\mu$ can be considered small. This could certainly be a too rough approximation, especially for the misaligned case, for which lines with footprint at the same magnetic co-latitude may have very different geometries (e.g., crossing or not the light cylinder, being more or less twisted, etc.).
Moreover, as mentioned before, our earlier spectral fitting studies showed that the required effective size of the accelerating region (i.e., the region along which most of the detected emission comes from) is of the order of $\lambda_0/R_{lc} \sim 10^{-3}$ or smaller. Therefore, one may think that the traversal width is small as well.
This limitation may be subject to further exploration in the future, if need arise. For the moment, this simplification is mostly motivated by the pursued effective approach with a limited number of free parameters.

\subsubsection{Co-rotating frame (CRF)}

We then consider the inclination angle $\Psi_\Omega$, i.e. the angle between the rotational axis, $\vec{\Omega}$, and the magnetic axis, $\vec{\mu}$. 
This will define the co-rotating reference frame (CRF), in which the star and the magnetic field lines do not rotate.
Fig. \ref{geo2} (bottom panel) is a useful reference for what follows. 

Since the $\vec{\Omega}-\vec{\mu}$ plane is described by $y_\mu=0$ in the MRF, then a rotation around the $y$ axis by $\Psi_\Omega$ will account for the meridional inclination:
\begin{equation}
 \hat{r}_\Omega = {\bf R}^y(-\Psi_\Omega) \hat{r}_\mu ~.
\end{equation}
Now, the rotational axis coincide with $\hat{z}_\Omega$ and the phase is $\phi_\Omega=0$ corresponds to the direction of the observer at the time reference $t=0$ (which is arbitrary, so that different choices produce a shift in phase)\footnote{In the extreme (and not interesting) case for which the observer is located along the rotational axis, $\theta_\Omega=0$ or $\pi$, the map becomes singular since the phase is ill-defined and has no meaning (a light curve constant in phase is seen by definition).}. In the CRF, the star is still and not rotating, a fixed observer at infinity draws a circle at a given co-latitude (calculated from the rotational axis) $\theta_{\Omega}$. Seen by an observer at infinity, the azimuthal coordinates directly corresponds to the rotational phase of the star.
%

\subsection{Time delay}

There is a further geometrical effect to consider when we calculate the radiation map: the phase flight delay $\xi_f$, which arises from the difference in the path length from the emission point to the observer. 
As a matter of fact, the radiation emitted at two different positions along the same line, with physical positions $\vec{r}_1$ and $\vec{r}_2$, will have a relative time delay given by 
\begin{equation}
\delta t_f = \frac{1}{c}(\vec{r}_1 - \vec{r}_2)\cdot\hat{r}_e~, 
\label{scalardelay}
\end{equation}
where $\hat{r}_e = \hat{r}_e(\lambda,\alpha,\chi)$ indicates the direction of emission, as given by the angles in Eqs.~(\ref{eq:angles_cone}).
%

The phase delay of the radiation emitted from a given $\vec{r}(\lambda)$ in the direction $\hat{r}_e$, is therefore $\phi_f(\lambda,\chi) = \Omega \delta t_f(\lambda) = \delta t_f c/R_{lc}$:
\begin{equation}\label{eq:phase_delay}
 \xi_f(\lambda,\chi) = \frac{1}{R_{lc}}\vec{r}\cdot\hat{r}_e + \xi_0~,
\end{equation}
where $\xi_0$ is the reference phase corresponding to $\phi_\Omega=0$, that we can set to zero without loss of generality.
Then, taking into account the radiation flight delay effect implies making an additional rotation to obtain the direction in the final sky map: 
\begin{equation}
 \hat{r}_{\rm rad} = {\bf R}^z(-\xi_f) \hat{r}_\Omega ~.
\end{equation}
The phase delay depends on the position along the line. If we are considering different lines, particles at the same $\lambda$ but on different lines will have a different phase delay.

\subsection{Aberration}

As pointed out by \cite{bai10}, aberration can play an important role in defining some details of the light curves, especially related to the cusps. 
This effect comes from special relativity, according to which the values of the magnetic field components change with the reference frame under a Lorentz transformation. 

In particular, in presence of a co-rotation velocity of the lines, the magnetic field seen by the observer is different to the one seen in the co-rotating frame, thus the direction of photon emission is deviated. For instance, a purely rotating poloidal field is reduced by a Lorentz factor $\Gamma$, and a toroidal component $\propto \Gamma$ appears due to the co-rotation velocity, thus becoming non-negligible close to the light cylinder. 
When the dipole is not aligned, things are more complicated, but in general, in the inertial (observer) frame, a component of the magnetic field in the direction of the velocity appears to be larger than in the co-rotating frame.
In what \cite{bai10} called {\it instantaneous co-rotating frame}, which is what we have called LRF, 
the poloidal magnetic field is reduced by a factor $\Gamma$ (see \cite{bai10} for the details about the Lorentz transformations).
Therefore, as shown in detail by \cite{bai10}, when we go back to the observer frame, the directions of the photons are aberrated, and their toroidal and poloidal emission directions (where the poloidal/toroidal decomposition is made with respect to the rotation axis) are given by
\begin{eqnarray}
 \hat{r}_{e,t} & = & \frac{\beta_{\rm rot} B_p^2 \pm B_t B_0'}{B_0^2} ~, \nonumber \\
 \hat{r}_{e,p} & = & \frac{\pm B_0' -\beta_{\rm rot}B_t}{B_0^2}\vec{B}_p ~ ,
\end{eqnarray}
where $B_0^{'2} = B_t^2 + (1 - \beta_{\rm rot}^2)B_p^2$ and $B_0^2 = B_t^2 + B_p^2$.
For instance, in the limit of $B_t = 0$ (e.g., a dipole aligned with the rotational axis), this means that the aberration would introduce a component in the toroidal direction (e.g., in the azimuthal direction around the rotation axis):
\begin{eqnarray}
\hat{r}_{e,t} & = & \beta_{\rm rot} ~, \nonumber \\
\hat{r}_{e,p} & = & \sqrt{1 - \beta_{\rm rot}^2}\frac{\vec{B}_p}{|\vec{B}_p|} ~.
\end{eqnarray}
In the  limit of $\beta_{\rm rot}=0$, then $\hat{r}_{e,t} \rightarrow B_t/B_0$, and $\hat{r}_{e,p} = {\vec{B}_p}/{|\vec{B}_p|}$, which is what we are already using $\hat{r}_e = \vec{B}/B_0$.
Overall, knowing a functional form for $\beta_{\rm rot}(\lambda)$ and the exact form of the magnetic field components $B_t$ and $\vec{B}_p$, we can calculate the aberration given by the velocity.

However, we pose that introducing this level of refinement in the calculation in our approach would likely bring in more problems than advantages, and can be unwarranted. 
First of all, it would mean to introduce free functional forms, which are not trivial. 
Second, and more important, \cite{bai10} --as well as other authors of similar works-- carefully model the geometry of a rotating retarded vacuum dipole, assuming at the same time $\beta_{\rm rot} = \varpi/R_{lc}$, where $\varpi$ is the distance from the rotation axis. However, this description clearly breaks down at the light cylinder, since the co-rotating velocity reaches the speed of light.

As a matter of fact, retarded vacuum dipoles are not thought to hold as a good description for rotating pulsars, especially in the outer magnetosphere where the lines get more distorted. 
That approach does not allow to have emission regions beyond the light cylinder.
But indeed, accelerating regions are supposed to appear either in the current sheet outside the light cylinder (where then the dipole description is incorrect), or in the border among the open and field close lines, where the opening out of the lines is an important effect, being connected to what is known as the wind zone. There are many numerical works that show the form of the magnetosphere in these regions, with force-free MHD (see for instance \cite{contopoulos99,gruzinov06,mckinney06,spitkovsky06,carrasco18}) or particle-in-cell \citep{kalapotharakos18} simulations.

In order to use a more consistent rotating dipole, then, one should rely on a limited number of numerical solutions, for a limited set of inclination parameters. 
Those configurations are not analytically describable, therefore limiting the exploration only
to the few cases  available from simulations (with fixed light cylinder, usually very close to the surface, fixed inclination angle, etc.). 
Thus, whereas it is true that the accuracy of the Lorentz transformation, based on the perfect co-rotation approximation, would represent an important inclusion in the model, changing from a vacuum dipole to a more realistic or more general configuration most likely will have much stronger effects on the light curves and spectra than aberration itself.

More importantly, we also note that even including the aberration effect, the changes are important only in the peaks of the light curves \cite{bai10}, and mostly 
goes without introducing qualitative differences in their overall shape. 
%
Because of that,
we shall not model aberration here, keeping it in mind as a possible further complication of the model.

\subsection{Final expression for the transformation}

The total coordinate transformation from the emission directions as seen in the PRF, $\hat{r}_e$ (the rightmost vector in the right hand side of the equation below, given by eq.~(\ref{eq:angles_cone})) to the Delayed Radiation Reference Frame (DRRF) is given by
\begin{equation}
\label{eq:total_rot}
\!\!\!\!\!\!\!\!
\hat{r}_{\rm map} = {\bf R}^z(-\xi_f){\bf R}^y(-\Psi_\Omega) {\bf R}^z(-\xi_\mu) {\bf R}^y(-\Psi_\mu) {\bf R}^y(-\Psi_t)~\hat{r}_e,
\end{equation}
where we remind that $\xi_f$ and $\Psi_t$ depend on $\lambda$. This relation allows to transform the coordinates of the emitted radiation direction $\hat{r}_e(\lambda,\chi)$ into the corresponding coordinates in the DRRF, $\hat{r}_{\rm map}(\lambda,\chi)$. Finally, the latter Cartesian coordinates can be translated into the usual angular ones, $(\theta_{\rm map},\phi_{\rm map})$, using Eqs.~(\ref{eq:theta_def})-(\ref{eq:phi_def}).

The variables in the transformation of eq.~(\ref{eq:total_rot}) play different roles. 
Some are free parameters while others are just
used in averaging integrals and to define the position along the trajectory of the particles. In particular, note that \\

(i) both the running variable along the trajectory, $\lambda \in [\lambda_{in},\lambda_{out}]$, and the gyration angle, $\chi \in [0,2\pi]$, enter in the coordinates of the emission in the LRF and in the time-delay effect (the last rotation to apply), and \\

(ii) $\Psi_\Omega$, $\xi_\mu$, $\Psi_\mu$ are three free parameters from which the radiation map will depend upon
 (in addition of the other parameters 
of the spectral-only model, and possibly fixed by the spectral analysis if observational data exist).

\section{Emission map}

In order to obtain the emission map, we start considering the direction of the particles.
Remember that, for a given position $\lambda$, each particle points toward a gyration-averaged direction given by Eqs.~(\ref{eq:angles_cone})-(\ref{eq:angles_cone_phi}) in the PRF.
We need to employ eq.~(\ref{eq:total_rot}) to find the corresponding values in the DRRF sky, $(\theta_{\rm map},\phi_{\rm map})$. Note that they depend on the trajectory-related variables, $\alpha(\lambda),\chi$, and on a set of three parametric angles: $\Psi_\mu,\xi_\mu,\Psi_\Omega$. We can formalize the {\it directions distribution} (i.e., per unit DRRF solid angle $d\Omega_\Omega = \sin\theta_\Omega d\theta_\Omega d\phi_\Omega$, normalized to unity) pointed by a particle at a given $\lambda$, averaged over the gyration angle $\chi$ (for the reasons mentioned in \S~\ref{sec:prf}), as:
\begin{equation}\label{eq:direction_distribution}
\frac{d D(\lambda)}{d\Omega_\Omega} = \frac{1}{2\pi\sin\theta_\Omega}\int_0^{2\pi} 
\delta[\theta_\Omega - \theta_{\rm map}(\chi)] \delta[\phi_\Omega - \phi_{\rm map}(\chi)] d\chi~,
\end{equation}
where we left implicit the dependencies of $\theta_{\rm map}$ and $\phi_{\rm map}$ on $\lambda$ and $\chi$, and $\delta[\cdot]\delta[\cdot]$ is the Dirac delta function in unit sphere coordinates, which is here formally used to describe the projection in the DRRF sky of the emission directions pointed by the particle during a gyration period. These directions are simply seen as a ring with radius $\alpha$ in the PRF, eqs.~(\ref{eq:angles_cone})-(\ref{eq:angles_cone_phi}), but the rotations have the effect to distort its shape in the DRRF sky (see also the benchmark tests, \S~\ref{sec:tests}).

We can now consider a generic {\it single-particle spectral photon flux} $I_E$, defined as the number of photons emitted per unit energy, per unit time, per particle, at a given point. The distribution in the DRRF map of the {\em single-particle spectral photon intensity}, defined as the gyration-averaged single-particle spectral flux per unit solid angle, is:
\begin{equation}\label{eq:distribution_cone}
\frac{d I_E}{d\Omega_\Omega} = I_E(\lambda) \frac{d D(\lambda)}{d\Omega_\Omega}~.
\end{equation}
If one assumes that $I_E$ is given by the synchro-curvature emission as derived in \cite{paper0}, then it will depend on the photon energy $E$, and on the values of $\alpha$, $r_c$ and $B$. They are ultimately reduced to a $\lambda$-dependency if the particle trajectory is solved as well for a specific pulsar of period $P$ and accelerating region properties ($E_\parallel, b$), where the latter is the magnetic gradient as defined in \cite{torres18} and references therein. In that case, the expression~(\ref{eq:distribution_cone}) depends on
$(E,\lambda,\Psi_\mu,\xi_\mu,\Psi_\Omega)$.
The next step is to consider a particle distribution along the line, $d{\cal N}/d\lambda(\lambda)$. Furthermore, one can consider a finite size (and/or number) of accelerating region(s), describing the corresponding bundle of lines by the a range of value for $\Psi_\mu$ and $\xi_\mu$ (for the magnetic meridional and azimuthal extensions, respectively).
 
The convolution of eq.~(\ref{eq:distribution_cone}) with the particle distribution, integrated over the traversal size of the accelerating region and along the line, gives the final expression of the total {\em spectral photon intensity map} ({\it emission map} hereafter for simplicity) defined as the number of photons emitted, per unit solid angle, per unit time, per unit energy:
\begin{equation}\label{eq:map_final}
M_E(\theta_\Omega,\phi_\Omega) = \int_{\lambda_{\rm in}}^{\lambda^{\rm out}} \int_{\delta\xi_\mu}\int_{\delta\Psi_\mu}\frac{d{\cal N} }{d\lambda}\frac{d I_E}{d\Omega_\Omega}  d\Psi_\mu d\xi_\mu d\lambda ~.
\end{equation}
This map represents the energy-dependent photon distribution emitted by the whole system over the unit sphere, already corrected for the time of flight 
(and for aberration if included in the calculation). Remember that, in the DRRF coordinates, $\theta_\Omega$ sets the observer's line of sight ($\theta_\Omega=0$ coincident the direction of the rotation vector), while $\phi_\Omega$ spans the spin phases.

In order to reduce the number of free parameters at play, we shall assume that, in the case of a wide region (finite values of $\delta\Psi_\mu$ and/or $\delta\xi_\mu$), the trajectories and the particle distributions are roughly the same for all the lines considered (as mentioned in \S~\ref{sec:mrf}).
Another simplification is to assume $d{\cal N}/d\lambda(\lambda)$ constant along a magnetic field line, which would effectively hold if the pair production in the region itself is not dominating the total number of particles being accelerated. These assumptions follow the same effective approach of the previous works, to keep the number of free parameters limited. They could be relaxed, introducing further complications in the model.

Last, we note that we can directly recover the effective value ${d{\cal N_{\rm eff}}}/{d\lambda}$ used in previous works where we performed spectral fits without any geometry or emission map calculation.\footnote{In the notation of \cite{paper0} to \cite{torres18} we use ${dN}/{dx}$.} 
That value represented the number of particles per unit $\lambda$ which effectively emit towards an observer at infinity located at a direction $\theta_\Omega$, and is given by the integration in phase and accelerating region transversal sizes of the directions distribution, eq.~(\ref{eq:direction_distribution}):
\begin{equation}\label{eq:map_dir}
\frac{d{\cal N_{\rm eff}}}{d\lambda}(\theta_\Omega,\lambda) =  \frac{d{\cal N} }{d\lambda} \int_{0}^{2\pi} \int_{\delta\xi_\mu}\int_{\delta\Psi_\mu}\frac{d D(\lambda)}{d\Omega_\Omega}  d\Psi_\mu d\xi_\mu d\phi_\Omega ~.
\end{equation}
This is the key to directly connect light curves and spectra, on which we shall focus our forthcoming work.

\section{Numerical procedure}\label{sec:method}

In order to numerically obtain the emission map in the DRRF, from where the light curves can be extracted, we shall apply the following numerical approach. 
\begin{enumerate}
	
	\item Set the magnetic line curvature dependence ($r_c(\lambda)$), the magnetic field dependence along the line ($B(\lambda)$), the initial pitch angle (see \citet{paper3, torres18} for details), and assume a given particle distribution ${d{\cal N} }/{d\lambda}$ (which we shall just take as a constant, see below).\\
	
	\item Consider a set of $\bar N_\lambda$ values $\lambda_i$, $i \in [0,\bar N_\lambda-1]$, parameterizing the line.
	At each $\lambda_i$, consider the corresponding $B$ and $r_c$ and calculate the trajectories of the particle by numerically solving the equations of motion. This provides the corresponding values of the pitch angle $\alpha_i$ and the Lorentz factor $\Gamma_i$.\\
	
	\item By solving the Frenet-Serret equations, we obtain the tangent and normal directions of the magnetic line in the PRF $\hat{t}_i$, $\hat{n}_i$. Given a physical distance from the inner part of the accelerating region to the star's center, $r_{\rm in}$, we can then calculate the set of physical positions $\vec{r}_i$, necessary to determine the relative phase delay between the emission at different positions. At this step, the trajectory of the particles and the geometry of the line are completely defined.\\	
	
	\item Considering each position $\lambda_i$, calculate the contribution to the single-particle spectral photon flux $I_i=I_E(\lambda_i)d\lambda_i$ (where $d\lambda_i$ is the discrete numerical bin around the considered position) for the given effective parameters (accelerating electric field $E_{||}$, contrast $x_0/R_{lc}$, magnetic gradient $b$, and normalization $N_0$) that define the average spectral energy distribution. This step was the key in our earlier spectral-only studies, e.g. \citet{torres18}.  
	\\

	\item At each position along the line, the emission occurs in the projected circle with opening $\alpha$ and center $(\theta_t,0)$ in the LRF. Therefore, for every $\theta_{t,i}(\lambda_i)$, we define a set of $\bar N_\chi$, equally spaced discrete values of the angle $\chi_j \in [0,2\pi]$. For a given $\lambda_i$, we therefore consider that the corresponding emission is homogeneously spread across the discrete values $\theta_{e,ij}=\theta_{t,i} + \alpha_i\cos\chi_j$ and $\phi_{e,ij}= \alpha_i\sin\chi_j$. For every value of the pair of angles ($\theta_{e,ij},\phi_{e,ij}$), we compute also the corresponding Cartesian discrete coordinates in the LRF, $\hat{r}_{e,ij}$, via eq.~(\ref{eq:unit_sphere}).
	\\

	\item Parameterize the angles of the inner accelerating region in the MRF: 
	$\bar N_l$ values of $\xi_\mu$ (the magnetic longitude), and $\bar N_k$ values of $\Psi_\mu$ (the magnetic co-latitude). 
	In principle, such values depend both on $k,l$, since moving the line the inner accelerating region can vary both in longitude and co-latitude: 
	we have discrete values given by $\xi_{\mu,kl}$ and $\Psi_{\mu,kl}$.
	In the simplest case, we consider only one line ($\bar N_l=\bar N_k=1$). Otherwise, we will have a uniform sampling of both angles, with bins  $d\xi_{\mu, kl}$ and $d\Psi_{\mu,kl}$.
	\\

	\item For each line, characterized by the indices ${l,k}$, we apply a set of rotations to each direction of emission $\hat{r}_{e,ijlk}$, corresponding to the emission direction in LRF (trajectory position $i$ and the sampling of angle $j$ in the circle of angular radius 
	given by the pitch angle $\alpha_i$), to obtain the coordinates in the map
	$\hat{r}_{{\rm map},ijkl} = {\bf R}^z(-\xi_f){\bf R}^y(-\Psi_\Omega) {\bf R}^z(-\xi_{\mu,kl}) {\bf R}^y(-\Psi_{\mu,kl}) \hat{r}_{e,ijkl}$. Then we transform the DRRF Cartesian coordinates, $\hat{r}_{{\rm map},ijlk}$, into the DRRF sky coordinates, ($\theta_{{\rm map},ijkl},\phi_{{\rm map},ijkl}$).
	\\

	\item Considering the DRRF sky (described by the coordinates $(\theta_\Omega,\phi_\Omega)$), sample it with a number of patches centered in $(\theta_p,\phi_q)$, with $p\in [0,\bar N_p]$, $q \in [0,2 \bar N_p]$ equally spaced in the ranges $\theta \in [0,\pi]$ and $\phi \in [0,2\pi]$, respectively. \\

	\item For each discrete emission direction numerically considered, $(\theta_{{\rm map},ijl},\phi_{{\rm map},ijlk})$ we find the patch of the DRRF sky wherein falls, identified by $(p,q)$, and add to it the emission given, as said above, by the value $I_E(\lambda_i) d\lambda_i d\chi_j$. We obtain the final emission map considering the discretized version of eq.~(\ref{eq:map_final}), where the integral is numerically given by the sum of $\bar{N}_{e,tot} = \bar N_\lambda\times \bar N_\chi \times \bar N_k\times \bar N_l$ contributions (i.e. for the different $\lambda_i$, $\chi_{ij}$, $\xi_{\mu,kl}$ and $\Psi_{\mu,kl}$):
	\begin{equation}
	M_\Omega (p,q) = \sum^*_{i,j,k,l} \left[\frac{d{\cal N} }{d\lambda}(\lambda_i)I_E(\lambda_i) ~d\lambda_i ~d\chi~ d\xi_\mu~ d\Psi_\mu \right]~,
	\end{equation}
	where the sum $\sum^*$ is performed only over those set of values of the indexes such that $(\theta_{{\rm map},ijlk},\phi_{{\rm map},ijlk})\in (p,q)$, $d\lambda_i$, $d\xi_\mu$,  $d\Psi_\mu$ take into account the discretized bins of the accelerating region volume, and $d\chi$ the uniform discretized bins of the gyration angle.
	\\

\end{enumerate}

The calculation of the emission over the discrete $2\bar N_p^2$ patches is performed by integrating the emission over a four-dimensional space parameters ($\lambda,\chi,\xi_\mu,\Psi_\mu$). 
We deal with the election of $\bar N_p, \bar N_\lambda, \bar N_\chi, \bar N_l, \bar N_k$. 
In order to save computational time, we consider a single direction (the tangent to the trajectory) instead of integrating over $\chi$, in case $\sin\alpha \ll \pi/\bar{N}_p$ (i.e., the circle of emission in the sky is very likely contained in the same patch, of angular opening $\sim\pi/\bar{N}_p \gg \alpha$). 
Furthermore keep in mind that some calculations are directly simplified if one takes the assumptions described at the end of the previous section: constant 
${d{\cal N} }/{d\lambda}$, single value of $\Psi_\mu$ and/or $\xi_\mu$.

We have performed convergence tests, by comparing the maps coming out from the same geometrical set-up but different resolutions. For a one-line case, we found that acceptable values, for which the maps are smooth and show numerical convergence, are typically $\simeq \bar{N}_\chi \simeq \bar{N}_\lambda \gtrsim 100$, with $\bar N_p^2 \ll \bar N_\lambda \times \bar N_\chi$ (typically, $\bar N_p \simeq 100$), in order to avoid an oversampling of the displayed sky patches (see Appendix). 
In other words, we want each sky patch to have enough statistics, which in our case are the number of discrete directions considered. 
Although the precise number can depend on the chosen geometry, and on the need to resolve fine details in the light curve, the recipe above can be considered as a rule-of-thumb.

\section{Benchmark tests}\label{sec:tests}

Hereafter we provide a set of tests with simplified geometries, in order to check the correct geometrical implementation. We will consider a given set of values for 
 $\Psi_\Omega$, 
 $\Psi_\mu$, and 
 $\xi_\mu$, which are the three main geometrical free parameters. Only for the sake of numerical testing some of the cases include a very wide range of $\xi_\mu$ (likely unphysical), keeping the same $\Psi_\mu$.

For these tests, we do not solve the trajectories of the particles, instead we shall first assume an ad hoc,
decreasing triviality of the parameters involved.
We shall also consider an energy-independent spectral photon flux $I_E=1$. 
These cases are not meant to represent realistic cases (which will be done elsewhere), 
but, and thanks to their relatively simpler geometry, allow us to test the overall implementation of the method.
Table~\ref{tab:models} shows the models tested.

\subsection{Geometrical tests for aligned rotators}

\begin{figure*}
	\centering
	\includegraphics[width=.45\textwidth]{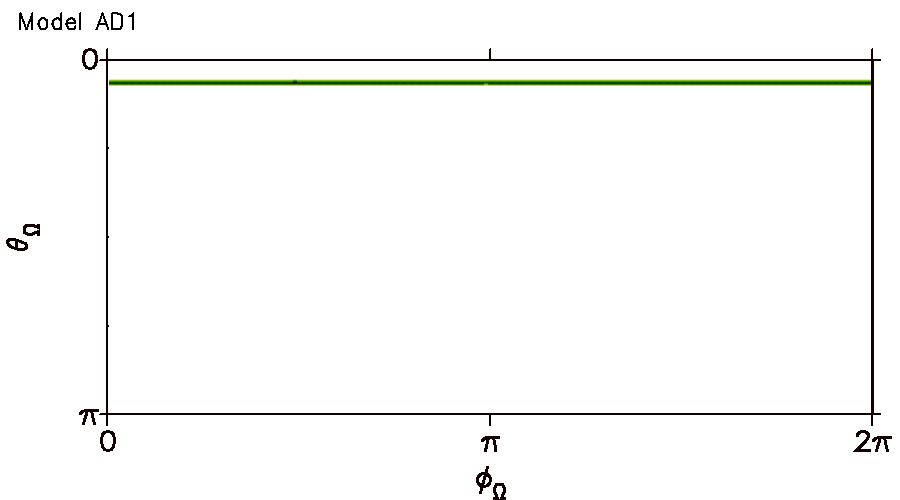}
	\includegraphics[width=.45\textwidth]{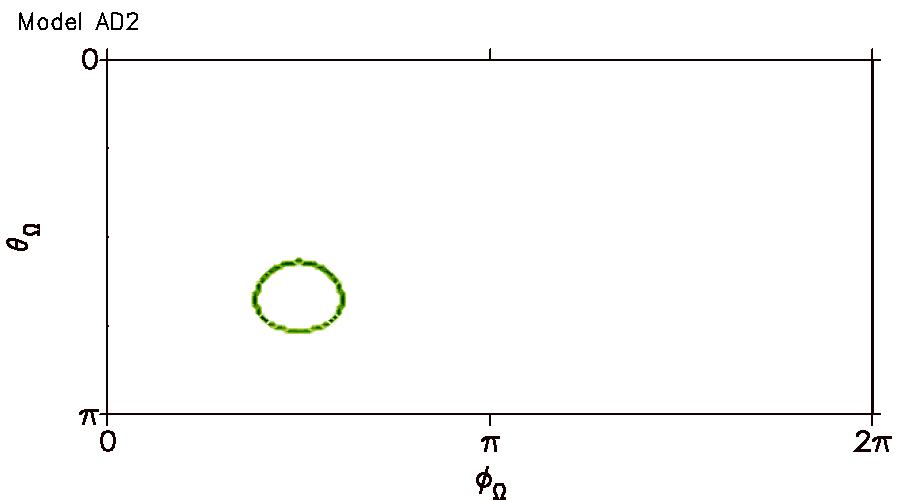}\\
	\vspace{0.2cm}
	\includegraphics[width=.45\textwidth]{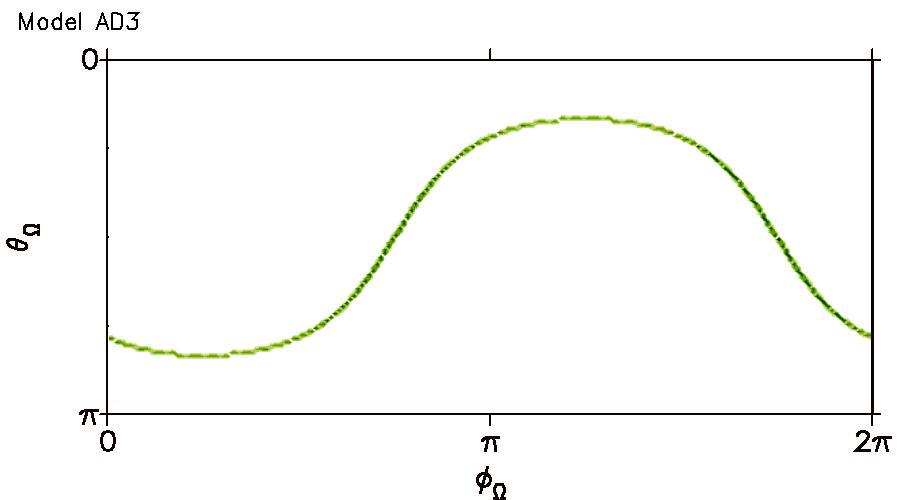}
	\includegraphics[width=.45\textwidth]{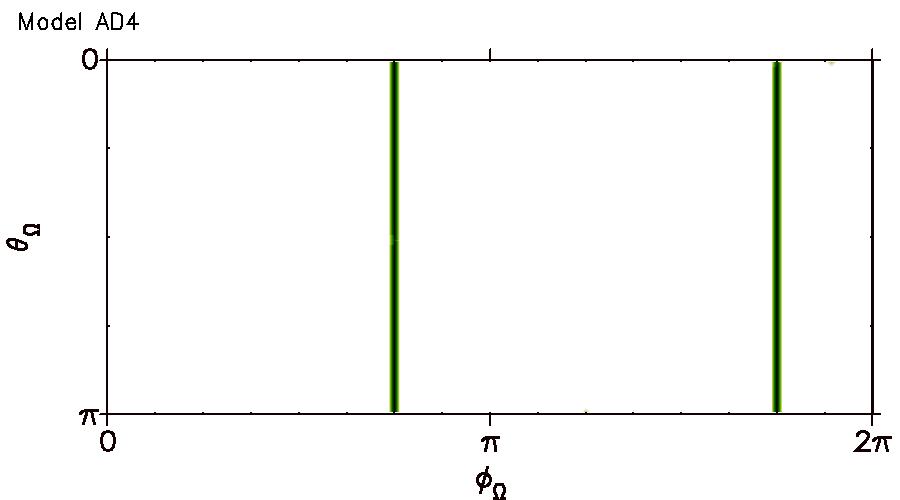}\\
	\vspace{0.2cm}
	\includegraphics[width=.45\textwidth]{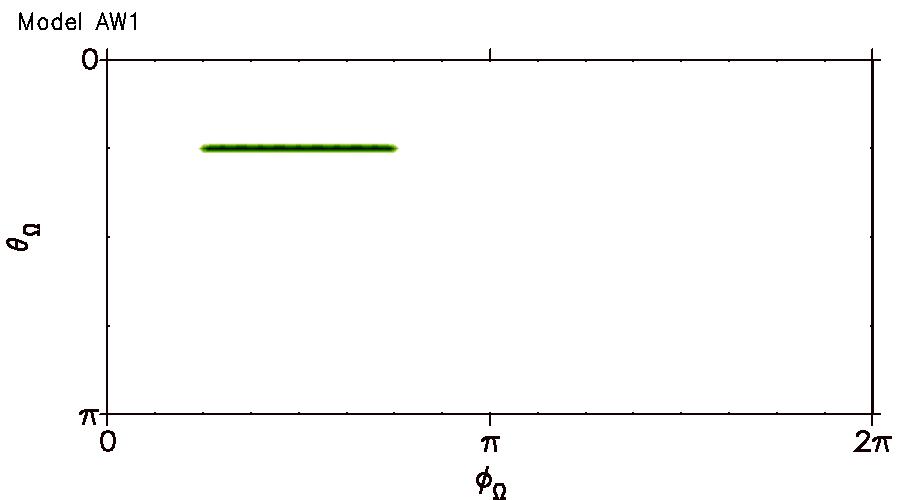}
	\includegraphics[width=.45\textwidth]{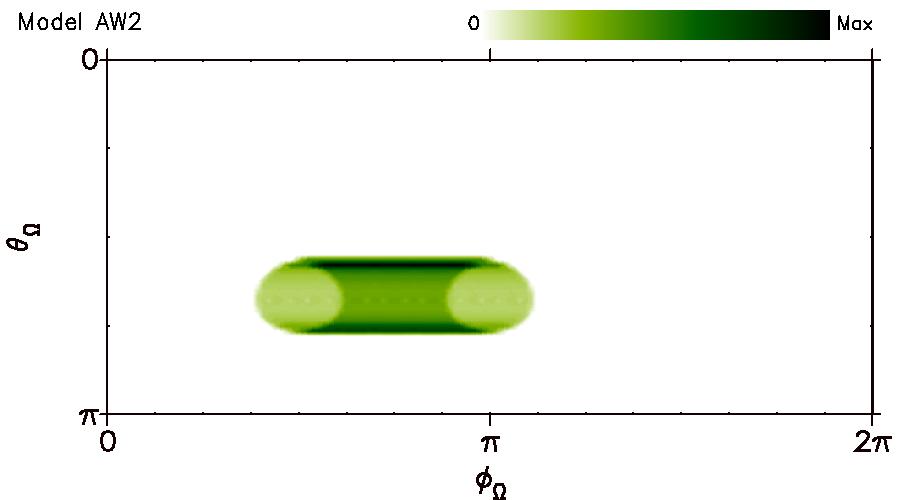}\\
	\vspace{0.2cm}
	\includegraphics[width=.45\textwidth]{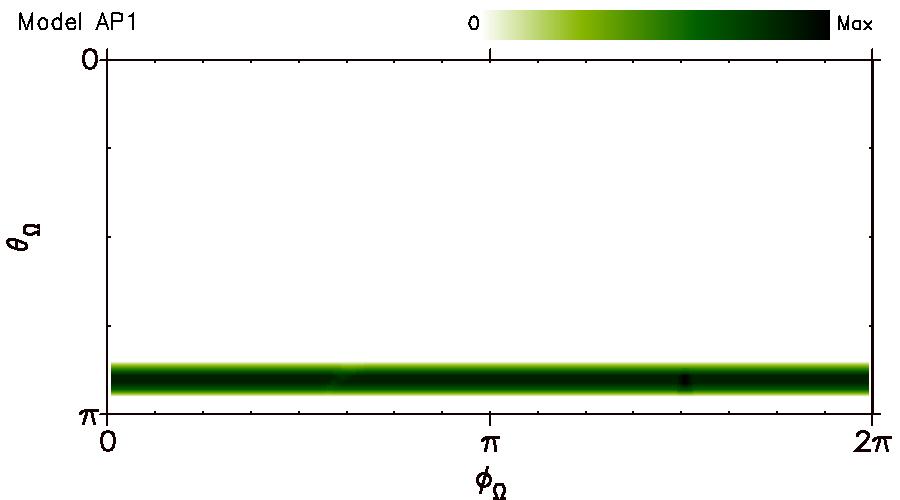}
	\includegraphics[width=.45\textwidth]{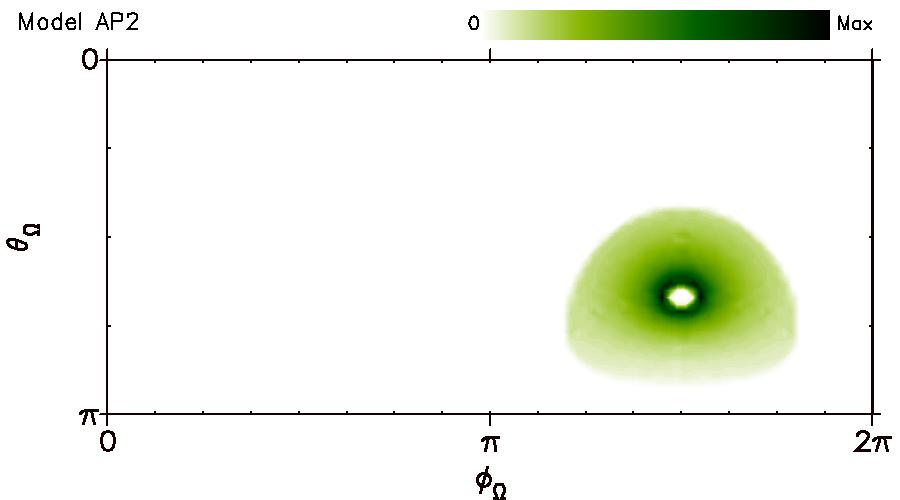}
	\caption{Emission maps for the aligned rotator geometrical models Axx described in Table~\ref{tab:models}. See text for discussion.}
	\label{fig:test_geo1}
\end{figure*}

First, we present a set of  cases neglecting the time delay, so that the specific length of the accelerating region and the position of the inner accelerating region do not play any role. 
The models AD1--AP2 test mainly the parameters $\alpha,\xi_\mu,\Psi_\mu$, considering an aligned rotator, $\Psi_\Omega=0$. Results are shown in Fig.~\ref{fig:test_geo1}. As a first set of models, we consider straight radial lines ($z_t=1$), fixed $\alpha$ and a single line. We progressively complicate the scenario.

\begin{itemize}
\item In the simplest cases, see models AD1-AD4, the map is just a projection of the lines considered. Note that these models (constant $z_t,\alpha,I_E$, single values of $\Psi_\mu,\xi_\mu$) basically tests the direction distribution, eq.~(\ref{eq:direction_distribution})  (hence the D in the model names), not considering any variation along the magnetic field line. In the first case, AD1, the straight line points to the polar direction $\Psi_\mu=0$, so that the emission encircles the axis by definition, with a co-latitude set by the chosen value of $\alpha$ (the circle around the axis becomes a straight line in the rectangular projection). In the model AD2, and whenever $\alpha \ll \Psi_\mu$, with a given $\xi_\mu$, the emission map appears like a circle (distorted by the projection) around the corresponding point in the sky map. Considering a high value of $\alpha \gtrsim \Psi_\mu$ (model AD3, in which case $\alpha=\pi/2$, $\Psi_\mu=\pi/3$), the emission cone is still centered around one specific direction (since we have a single value $\xi_\mu=\pi/4$), but, due to the large value of the opening angle, it encircles one pole, appearing as a wavy pattern. The same model can be brought to the extreme of vertical lines for the case $\Psi_\mu=\alpha=\pi/2$ (model AD4), with $\xi_\mu$ playing, as in the other cases, the role of displacing the emission along the phase. \\

\item When we add a width, like in model AW1, where we have set $\sin\alpha=0$ for each $\lambda$, so that the line point to a non-zero co-latitude ($\pi/4$), and the azimuthal width of the accelerating region is evident in the width of the emission projection (in this case, $\xi_\mu \in \left[\frac{1}{2}-\frac{3}{2}\right]\pi$) along the phase direction. \\

\end{itemize}

All the examples so far are ignoring the variation of the pitch angle, which is unrealistic.
We do not expect it to be constant, at most one can entertain it to be zero from the beginning, but this would not be the result of a random injection of particles in the accelerating region, like for instance in the case of pair production. Thus, we explore non-zero variations of the pitch angle next, to see the response of our geometrical approach.

\begin{itemize}
\item If we let $\sin\alpha$ vary linearly with $\lambda$, ranging between 0.4 and 0.2 (model AP1), and 0.7 and 0.1 (model AP2, P is used to refer to the pitch angle variation), we can see how we obtain a thickness in the projected emission, corresponding to the superposition of decreasing opening angles (smaller circles in the map) for increasing $\lambda$. Both cases show the same features, with the difference of where the emission is centered (close to one pole, AP1, or far from it, as in AP2).\\

\end{itemize}

These models show an equatorially specular map, $M_\Omega(\theta,\phi)\rightarrow M_\Omega(\pi-\theta,\phi)$, under any meridional reversal of inclination angle, $\Psi_\Omega \rightarrow (\pi - \Psi_\Omega)$, or co-latitude of the line in the MRF, $\Psi_\mu \rightarrow (\pi - \Psi_\mu)$. Similarly, any model shows a phase shift in the map, $M_\Omega(\theta,\phi)\rightarrow M_\Omega(\theta,\phi+\delta)$, for any shift of the accelerating region line azimuth in the MRF, $\xi_\mu \rightarrow (\xi_\mu + \delta)$. 
This can be seen comparing, for instance, the emission maps of the models AD1 and AP1 (specular values of $\Psi_\mu$), and AD2 and AP2 (shift $\pi$ in $\xi_\mu$), respectively (the thickness of the lines is due to the above-mentioned different $\sin\alpha(\lambda)$ dependence being tested).

\begin{table}
	\caption{Table of the benchmark geometric models, with no time delay considered (therefore, the models are not sensitive to the position of the inner region $r_{\rm in}$, and the field line length $L$). Models A are aligned rotators ($\Psi_\mu=0$), while models O are oblique rotators. With ''any`` we indicate that the map is independent on the choice of that parameter. See text for further information. Note that $\Psi_\Omega$ is a unique value by definition, while for these test cases we consider that $\Psi_\mu$ and $\xi_\mu$ can be a fixed value or a range, and $\sin\alpha$ and $z_t$ can be either a fixed value, or an analytical function of $\lambda$.} 
	\begin{center}
		\begin{tabular}{l c c c c c}
			\hline
			\hline
			Model & $\Psi_\mu$ & $\xi_\mu$ & $\Psi_\Omega$ & $\sin\alpha(\lambda)$ & $z_t(\lambda)$ \\
			\hline
			\hline
			\vspace{0.2cm}
			AD1   & 0 & any & 0 & fixed 0.2 & fixed 1 \\
			\vspace{0.2cm}
			AD2   & $\frac{2}{3}\pi$ & $\frac{1}{2}\pi$ & 0 & fixed 0.3 & fixed 1\\
			\vspace{0.2cm}
			AD3   & $\frac{1}{3}\pi$ & $\frac{1}{4}\pi$ & 0 & fixed 1 & fixed 1\\
			\vspace{0.2cm}
			AD4   & $\frac{1}{2}\pi$ & $\frac{1}{4}\pi$ & 0 & fixed 1 & fixed 1\\
			\vspace{0.2cm}
			AW1   & $\frac{1}{4}\pi$ & $\left[\frac{1}{2}-\frac{3}{2}\right]\pi$ & 0 & fixed 0 & fixed 1 \\
			\vspace{0.2cm}
			AW2   & $\frac{2}{3}\pi$ & $\left[\frac{1}{2}-1\right]\pi$ & 0 & fixed 0.3 & fixed 1\\
			\vspace{0.2cm}
			AP1   & $\pi$ & any & 0 & $=0.4 - 0.2\frac{\lambda}{L}$ & fixed 1 \\
			\vspace{0.2cm}
			AP2   & $\frac{2}{3}\pi$ & $\frac{3}{2}\pi$ & 0 & $=0.7 - 0.6\frac{\lambda}{L}$ & fixed 1 \\
			\hline
			\vspace{0.2cm}
			OD1 & 0 & any & $\frac{1}{3}\pi$  & fixed 0.2 & fixed 1\\
			\vspace{0.2cm}
			OD2 & 0 & any & $\frac{1}{3}\pi$  & fixed 0.2 & fixed 0.6 \\
			\vspace{0.2cm}
			OW1 & $\frac{1}{4}\pi$ & $[0-2]\pi$ & $\frac{1}{3}\pi$  & fixed 0.2 & fixed 1 \\
			\vspace{0.2cm}
			OW2 & $\frac{1}{2}\pi$ & $[0-2]\pi$ & $\frac{1}{3}\pi$  & fixed 0.2 & fixed 1 \\
			\vspace{0.2cm}
			OW3 & $\frac{2}{3}\pi$ & $[0.5-1]\pi$ & $\frac{1}{3}\pi$  & fixed 0.2 & fixed 1 \\
			\vspace{0.2cm}
			OP1 & $\frac{2}{3}\pi$ & $[0.5-1]\pi$ & $\frac{1}{3}\pi$  & $=0.6-0.2\frac{\lambda}{L}$ &  fixed 1 \\
			\vspace{0.2cm}
			OP2 & $\frac{2}{3}\pi$ & $[0.5-1]\pi$ & $\frac{1}{3}\pi$  & $=0.6-0.6\frac{\lambda}{L}$ & fixed 1 \\			
			\vspace{0.2cm}
			OZ1 & $\frac{2}{3}\pi$ & $[0.5-1]\pi$ & $\frac{1}{3}\pi$  & $0.2$  fixed & $=1-\frac{\lambda}{L}$ \\
			\vspace{0.2cm}
			OZ2 & $\frac{2}{3}\pi$ & $[0.5-1]\pi$ & $\frac{1}{3}\pi$  & $0.2$  fixed & $=1-0.2\frac{\lambda}{L}$\\
			\vspace{0.2cm}
			OZP & $\frac{2}{3}\pi$ & $[0.5-1]\pi$ & $\frac{1}{3}\pi$  & $=0.2-0.2\frac{\lambda}{L}$ & $=1-0.2\frac{\lambda}{L}$\\
			\hline
			\hline
		\end{tabular}
	\end{center}
	\label{tab:models}
\end{table} 

\begin{figure*}
	\centering
	\includegraphics[width=.45\textwidth]{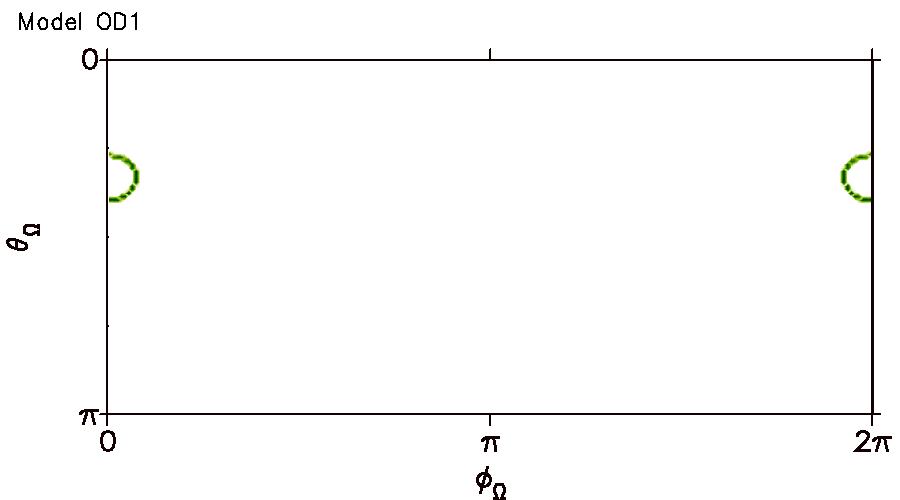}
	\includegraphics[width=.45\textwidth]{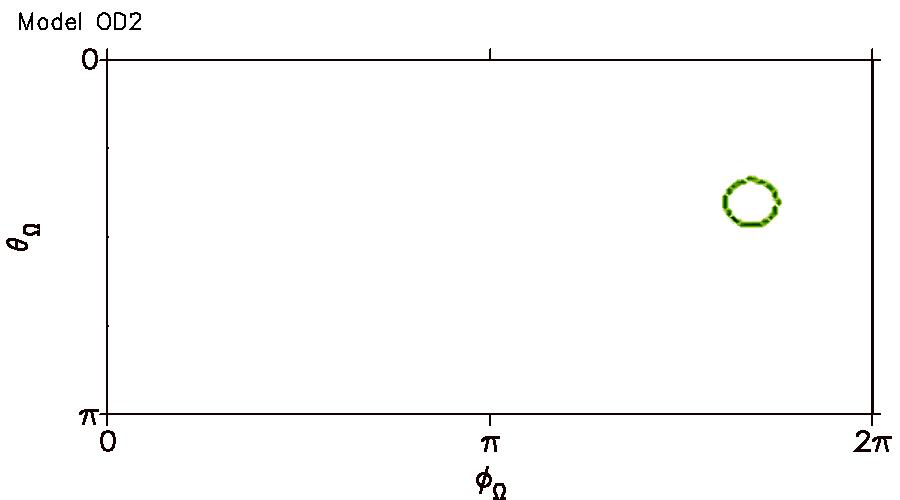}\\
	\includegraphics[width=.45\textwidth]{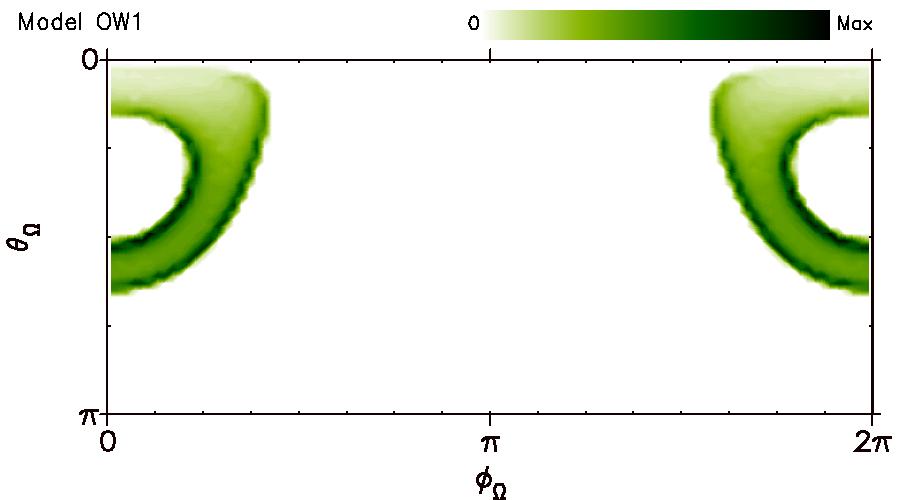}
	\includegraphics[width=.45\textwidth]{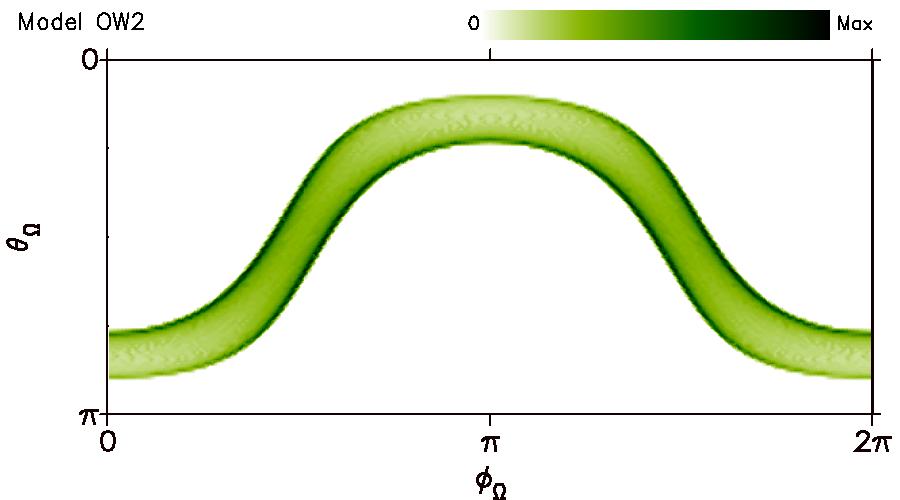}\\
	\includegraphics[width=.45\textwidth]{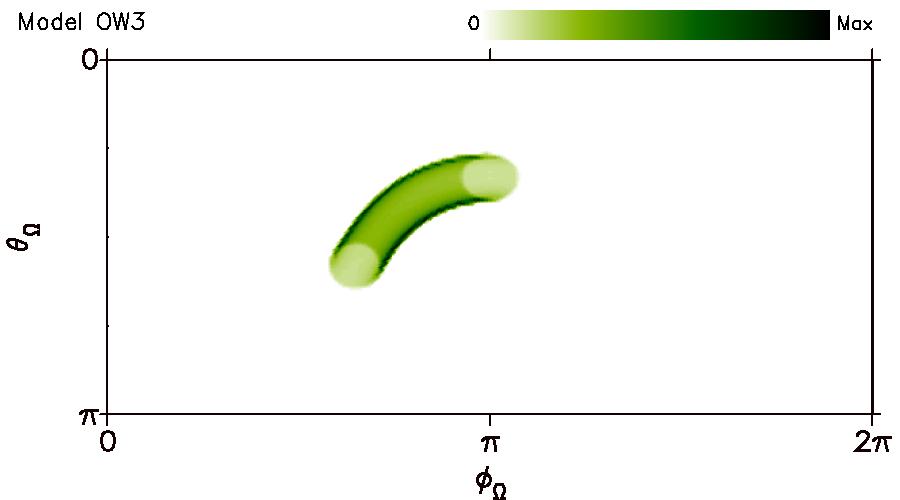}
	\includegraphics[width=.45\textwidth]{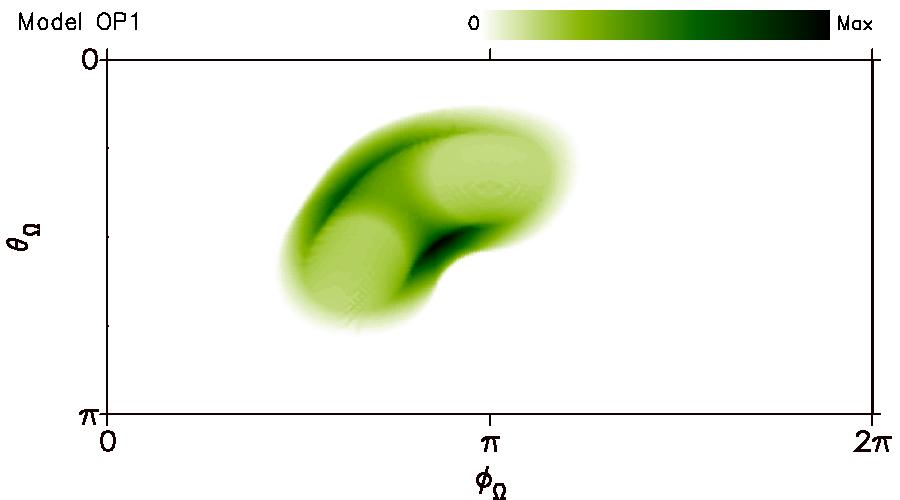}\\
	\includegraphics[width=.45\textwidth]{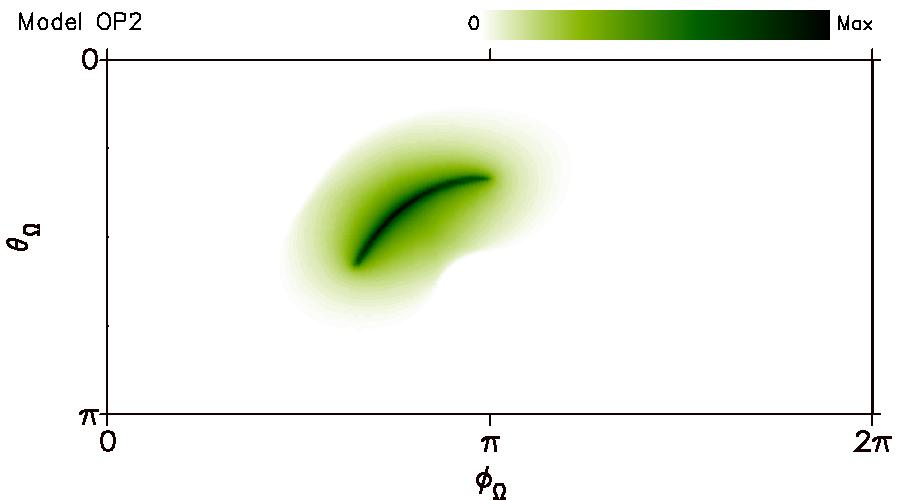}
	\includegraphics[width=.45\textwidth]{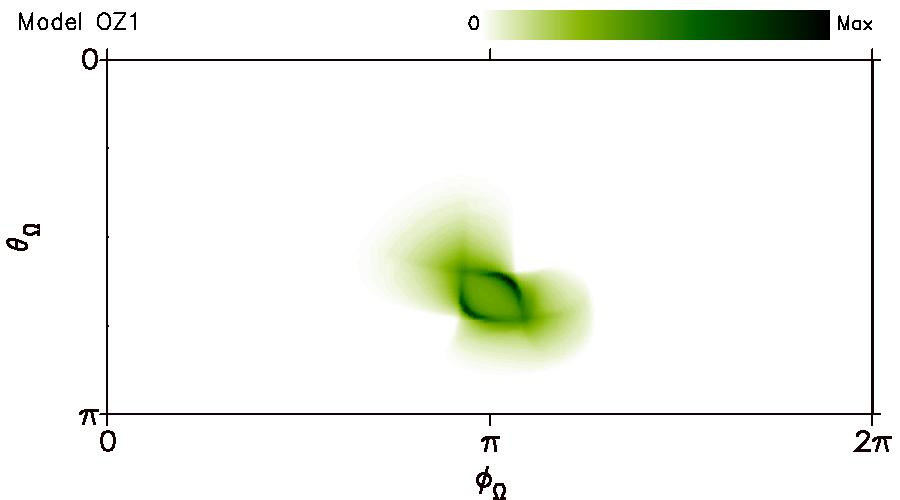}\\
	\includegraphics[width=.45\textwidth]{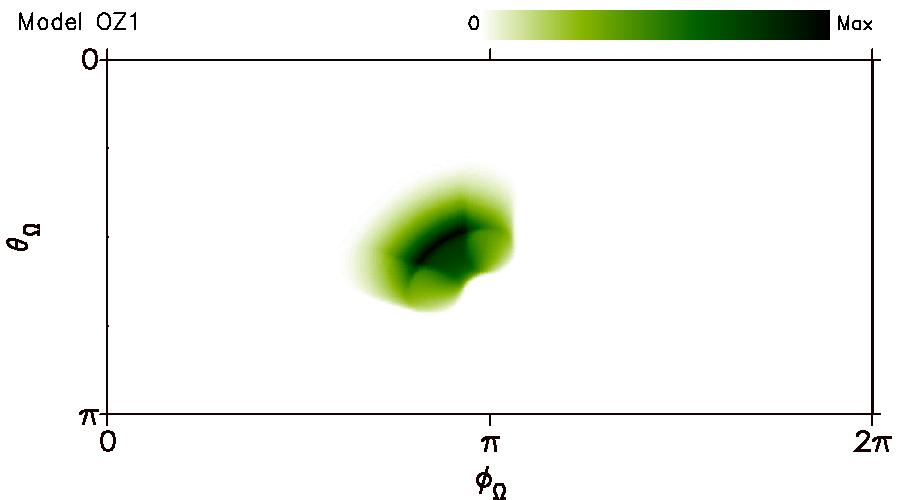}
	\includegraphics[width=.45\textwidth]{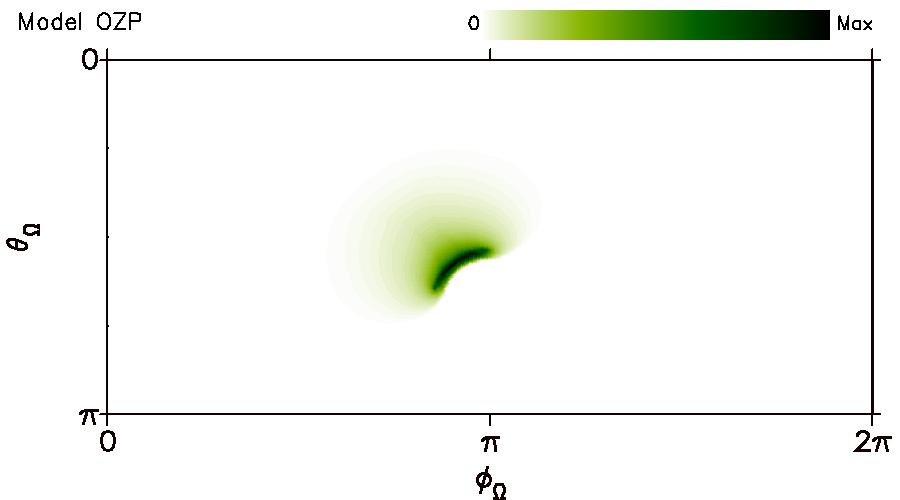}
	\caption{Emission maps for the oblique rotator geometrical models Oxx described in Table~\ref{tab:models}. See text for discussion.}
	\label{fig:test_geo2}
\end{figure*}


\begin{figure*}
	\centering
	\includegraphics[width=.45\textwidth]{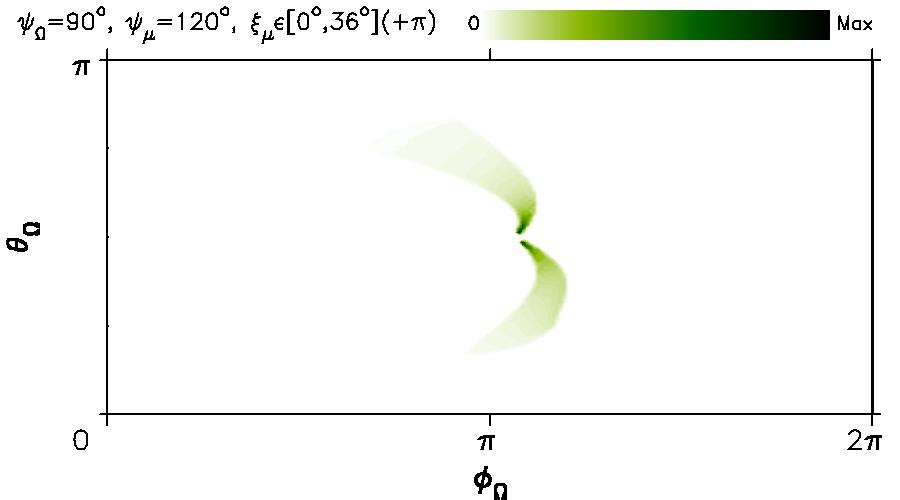}
	\includegraphics[width=.45\textwidth]{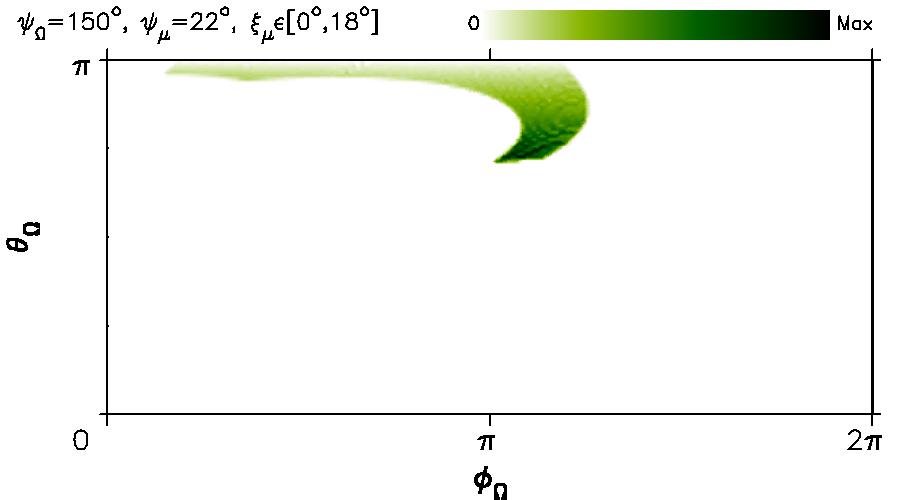}
	\includegraphics[width=.45\textwidth]{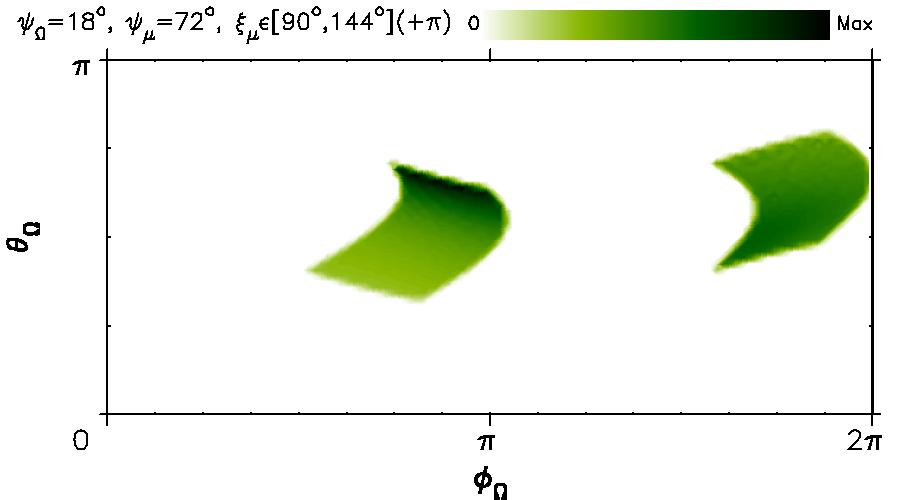}
	\includegraphics[width=.45\textwidth]{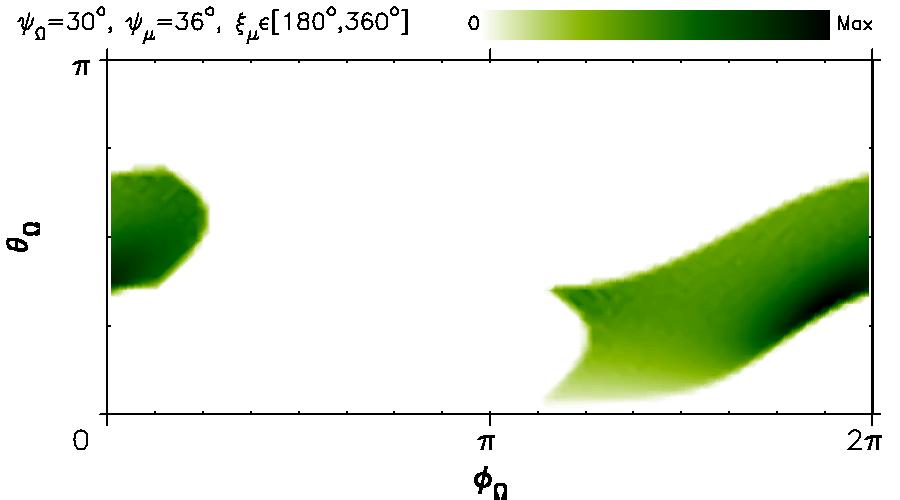}
	\caption{Emission maps from a pulsar-like, trajectory-based simplified model (see text for parameters). The different geometrical parameters are indicated in each panel, where the ($+\pi$) indicates that we consider a second range displaced by $\pi$ compared to the first one, indicated explicitly. We consider here $\bar N_p=100$. See text for discussion.}
	\label{fig:test_psr}
\end{figure*}

\begin{figure*}
	\centering
	\includegraphics[width=.45\textwidth]{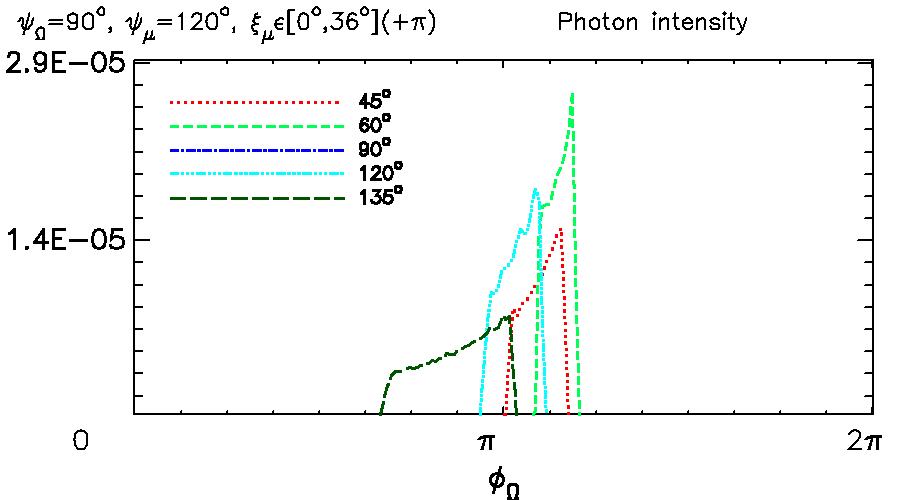}
	\includegraphics[width=.45\textwidth]{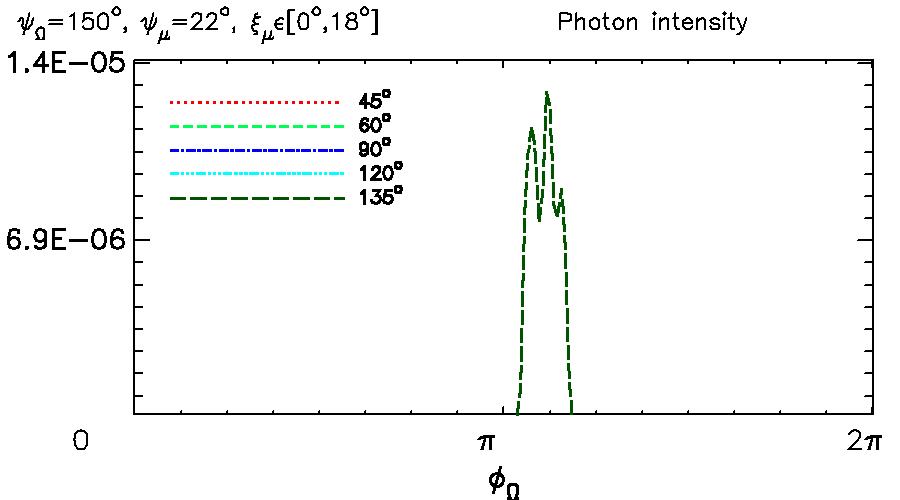}
	\includegraphics[width=.45\textwidth]{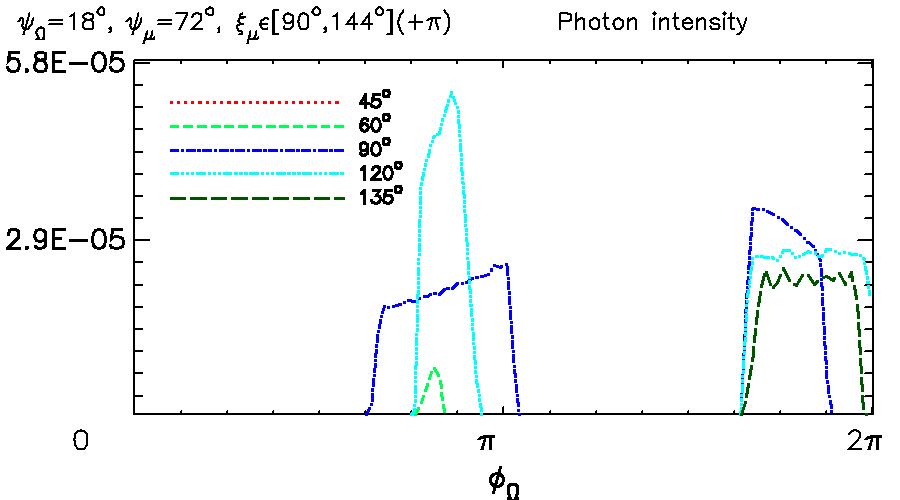}
	\includegraphics[width=.45\textwidth]{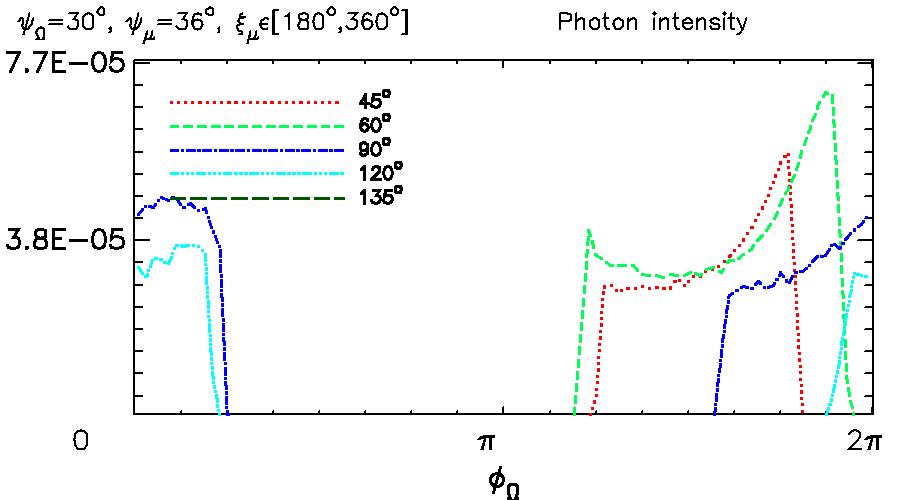}	\caption{Light curves seen by observers located at different $\theta_\Omega$ (indicated in the legend), obtained by cutting the emission maps shown in Fig.~\ref{fig:test_psr}.}
	\label{fig:test_psr_lc}
\end{figure*}

\subsection{Geometrical tests for oblique rotators}

We now relax the alignment hypothesis ($\Psi_\Omega=0$), and consider again different simplified geometries with an arbitrary value that we set to $\Psi_\Omega=\pi/3$. We describe the family of the models O, shown in Fig.~\ref{fig:test_geo2}.

\begin{itemize}
\item The first two cases (OD1 and OD2) consider the values $\Psi_\mu=0$, $\sin\alpha=0.2$, with the only difference between them found in the value of the tangent direction (in both cases, constant): $z_t=1$ in OD1 (as in the models A and the other models O), and $z_t=0.6$ in OD2. The first case corresponds to the emission along a cone centered around the magnetic axis, that in the sky map is defined at phase equal to 0. The second case, instead, moves the position according to the tangent direction chosen and the azimuthal position of the line considered. \\

\item In the models labelled OW1 and OW2, we consider an accelerating region that encompasses the star entirely in the azimuthal direction, and, with the same fixed value of $\sin\alpha=0.2$, explore the dependence on $\Psi_\mu$. The difference in the patterns appears because for Model OW1 $\Psi_\mu<\Psi_\Omega$, while for model OW2 $\Psi_\mu>\Psi_\Omega$, so that the wavy pattern appears again (for the same reason as it does in model AD3 above). In both cases, the thickness of the pattern is due to the value of $\sin\alpha$, while its evolution in phase reflects the azimuthal range $\Delta \xi_\mu=2\pi$. \\

\item The latter effect can be understood better in model OW3, which only differs from OW1 and OW2 in the values $\Delta \xi_\mu = 0.5\pi$ (hence, a quarter-circle only appears), and $\Psi_\mu=(2/3)\pi$ (affecting the position in the map). \\
\end{itemize}

In order to increase the complexity we finally introduce $\lambda$-dependencies in $\sin\alpha$ (models OP1, OP2), $z_t$ (models OZ1, OZ2), or in both of them (model OZP). While the effect of the former can be understood easily from the combined examples above,  the latter has the effect of moving the center of the emission along the map. The larger the variation, the more spread appears the emission, with the specific trend depending on the non-trivial geometry adopted.


\section{Light curves from pulsar-based particle trajectories}

We now consider a last example using particle trajectories. We express the the local values of magnetic field and radius of curvature as a function of $\lambda$, setting, respectively, $B(\lambda)=B_s(R_\star/(r_{\rm in} + \lambda))^b$ (with $b$ being the magnetic gradient, set here to $b=2.5$) and $r_c(\lambda)=R_{\rm lc}((r_{\rm in} + \lambda)/R_{\rm lc})^\eta$ (here set $\eta=0.5$), the distance of the inner part of the region from the star $r_{\rm in}=0.5~R_{lc}$, the region length (measured along the magnetic field line) $L=R_{lc}$. The initial pitch angle of particles is $\alpha_{in}=\pi/4$.

We consider typical parameters of standard pulsars: spin period $P=100$ ms and $B_s=10^{12}$ G. We consider a constant value of $E_\parallel=10^8$~V/m (a typical value from our previous works \citep{paper0,paper1,paper2,paper3,paper4,torres18}), to which we refer also for further details about the significance of these parameters, and the characteristic parameters along the trajectories.

With the chosen parameters, we solve the equations of motion, eq.~(\ref{motion}), and the Frenet-Serret equations, to determine the directions where the charged particles are radiating, as described in \S~\ref{sec:method}. The emission map depends on the evolution of the physical properties (like pitch angle, Lorentz factor, etc.) along the trajectory. The trajectories give also spatial position of the emitting particles at a given $\lambda$, needed to take into account the time delay, eq.~(\ref{eq:phase_delay}).

For the purposes of considering the geometrical effects (the main aim of this paper), we consider uniform photon intensity and uniform particle distribution $I_E=d{\cal N}/d\lambda =1$ (both independent on $\lambda$ and on the line considered). We look at the bolometric map (not considering any energy distribution). Note, therefore, that the units of the intensities are arbitrary.

As in the simpler tests above, we consider different geometrical angles, as indicated in the legends of Fig.~\ref{fig:test_psr}, where we show the bolometric emission maps obtained by different choices of parameters. In Fig.~\ref{fig:test_psr_lc} we show the corresponding light curves $M_E(\theta_{\rm obs},\phi_\Omega)$ as seen by different observers, i.e. values of $\theta_\Omega=\theta_{\rm obs}$. We consider a fixed value for $\Psi_\Omega$, $\Psi_\mu$ and a range for $\xi_\mu$. Depending on the range of value of the free parameters, the emission maps can change a lot. Maps are very sensitive to the range of $\xi_\mu$: the wider the latter, the more spread the map. Note also that the map clearly show when we consider two different ranges of $\xi_\mu$.

Note that light curves can have very different shapes and intensities, showing one (many of them) or two peaks (e.g., the red or green light curves in the top right panel), or being invisible at all. These qualitative behaviours, together with a sensitivity analysis, can be used to infer the position, size and numbers of such regions, which here, for simplicity, we consider to be only one with different azimuthal width.

Scanning  the influence that each of the parameters has on the light curves, we could see that what dominates the shapes 
shown in the examples are the geometrical parameters, the length of the region $L$ and the curvature radius $r_c(\lambda)$ (in our case, parametrized by $\eta$ and $r_{\rm in}$). 
Maps are instead quite insensitive to the value of $E_\parallel$, $B_s$ and $b$, which are instead important for spectral fits. This is a result of the fact that we are considering a constant spectral photon intensity, so that, in this simplified test, the only impact of the three parameters is reduced to slight changes of the pitch angle evolution in the initial synchrotron-dominated part of the trajectory. This also serves as a test of the implementation: we do not expect significant changes on the light curves 
being produced by the parameters which can be spectrally constrained.

Last, note that these tests shown in this paper are not meant to be the most realistic geometrical description of the accelerating regions: we are showing the capabilities of the numerical approach. It is not the purpose of this paper to explore the range of parameters and assess their more physically likely values, which deserves a dedicated study (in preparation). Instead, these tests show how our approach can provide a large variety of maps and light curves with realistic particle trajectories. 

In particular, the constant $I_E$ used here is an oversimplification: besides not considering the spectral distribution, the bolometric intensity emitted grows fast with the Lorentz factor, i.e. the outer parts of the trajectory will provide more photons. When self-consistently implemented with the proper trajectory-dependent spectral intensity $I_E(\lambda)$, already calculated in our previous works \citep{paper0,paper3,paper4,torres18}, these maps and light curves can be actually used in our reverse engineering approach, in order to infer the values of the parameters from the comparison with data.

\section{Concluding remarks}

This paper lays the foundation for a 
general approach for computing 
multi-frequency and simultaneous spectral and light curve predictions for pulsars.
At this stage, we have on purpose focused on describing the geometry, and our use of the Frenet-Serret equations to determine the magnetic lines in the magnetospheres.
We have provided a detailed numerical scheme that prescribes how the necessary geometry can be obtained. 
Our scheme is already prepared to deal with any spectral photon flux required, in particular, with the synchro-curvature radiation that so successfully described 
the spectral energy distribution of pulsars.
When such is incorporated, we shall have all the necessary ingredients to produce a simultaneous, multi-frequency prediction of both light curves and spectra. 

Some assumptions and caveats have still to be kept in mind.
For instance, it is sane to recall that our magnetic lines have no torsion (i.e., twist). This might end being not a realistic approximation, having in mind the sweepback of lines caused by rotation, especially strong near and beyond the light cylinder, as shown by numerical simulations (e.g., \cite{spitkovsky06} or any work about rotating magnetospheres).  The inclusion of torsion represent a further degree of freedom and can change the particle dynamics, and it is unclear whether it is indeed needed in case the relevant emission regions are as small as suggested in the spectral only models. The comparison with real light curves will tell whether this extra complexity is warranted or not.

A similar simplifying approximation is to assume a constant $\Psi_\mu$ for all lines, in case of a tranverally-extended accelerating region. This is certainly not exact for oblique rotators, if the region is large. However, again, this assumption can be overcome only at the cost of complicating the model and making it less effective in for fitting purposes, and again we are not yet uncertain such complication is warranted if the emission regions are small.

When a sizeable accelerating region is considered to have a finite width, we assume that the effective parameters describing it do not vary (i.e., the $r_c(\lambda)$ are the same for all lines). This is a simplification that has to be read in the perspective of having an effective and reduced parameterization of the complex magnetosphere and/or on the assumption that the width is not large so as to imply a large difference in $r_c(\lambda)$ or $E_{||}$ when moving across it.
 
 We consider that all particles are born (or injected) at the same place. That is, that a single trajectory describes the whole particle population. This is obviously not the case, and is done so (as was the case in our earlier papers) for a twofold reason. On the one hand, the obvious simplification that this brings make the model practical --it would be simply impossible to numerically consider the whole population of particles one by one. Some improvements can however be envisaged, as for instance, a distribution of injection places so that we average on a few trajectories before obtaining final results. However, on the other hand we noted already that the exact value of $\lambda_{in}$ is not dominating the spectral shape. On the contrary, it is actually the transition (the relative weight of the particle population) 
from synchrotron to curvature-dominated regimes what does dominate.

\section*{Acknowledgements}

DV acknowledges support from the Spanish Ministry of Economy, Industry and Competitiveness grants AYA2016-80289-P and AYA2017-82089-ERC (AEI/FEDER, UE). 
DFT similarly acknowledges support from grants PGC2018-095512-B-I00, SGR2017-1383, and AYA2017-92402-EXP. 

We thank Dr. Miguel Bezares for useful advises.

\appendix

\section{Numerical convergence}\label{app:convergence}

\begin{figure}
	\centering
	\includegraphics[width=.45\textwidth]{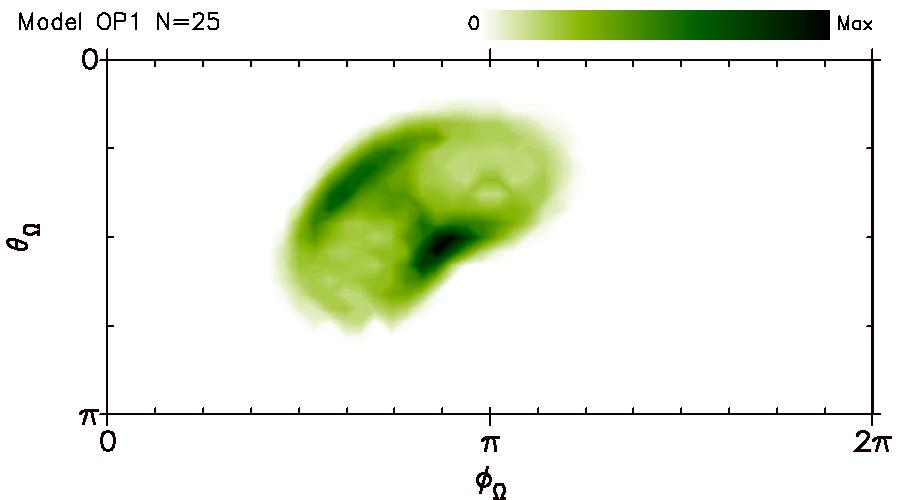}
	\includegraphics[width=.45\textwidth]{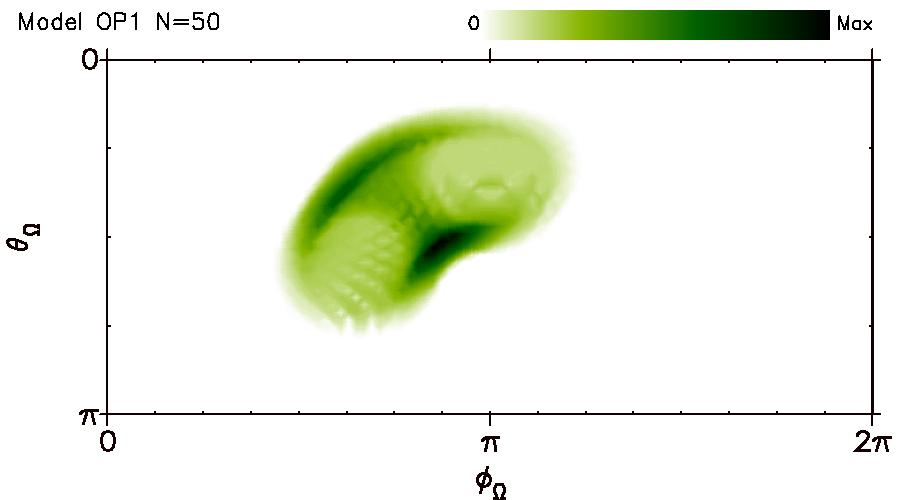}
	\includegraphics[width=.45\textwidth]{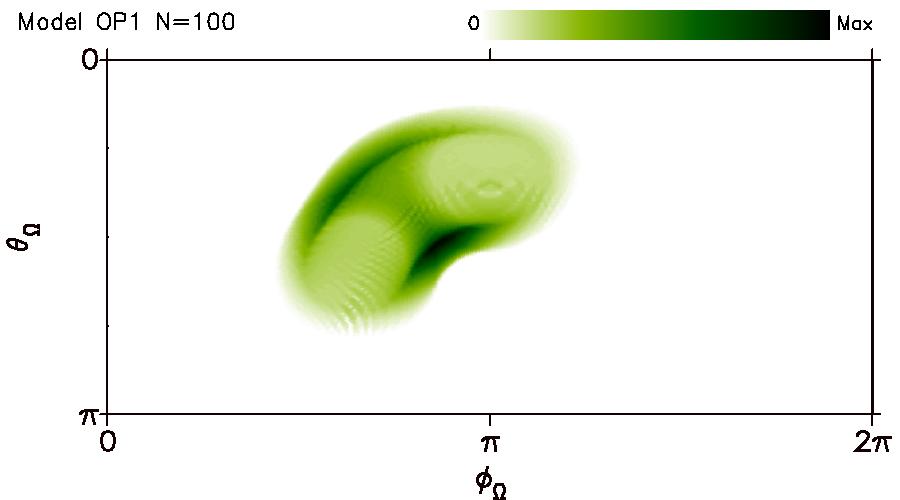}
	\includegraphics[width=.45\textwidth]{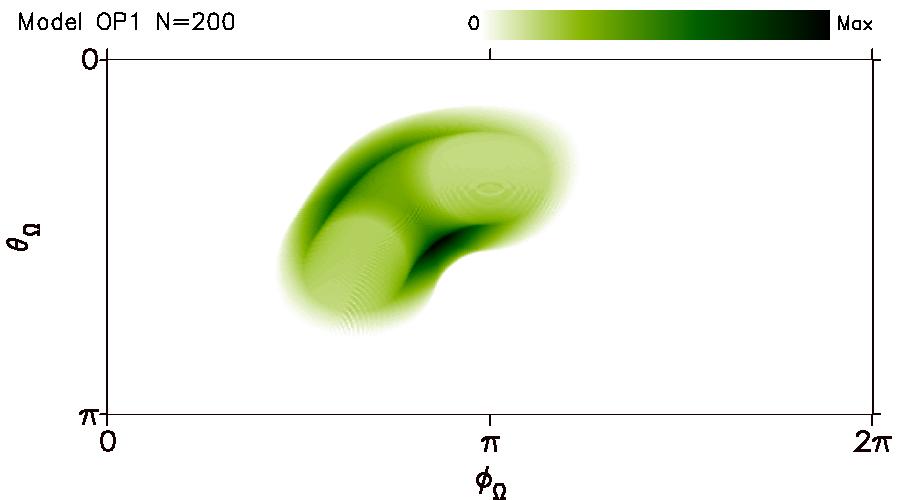}
	\caption{Convergence of the emission map for the model OP1, with $\bar N_p=\bar N_\lambda=\bar N_\chi=\bar N_l=25,50,100$ and 200 (from top to down). }
	\label{fig:test_geo_convergence}
\end{figure}

\begin{figure}
	\centering
	\includegraphics[width=.45\textwidth]{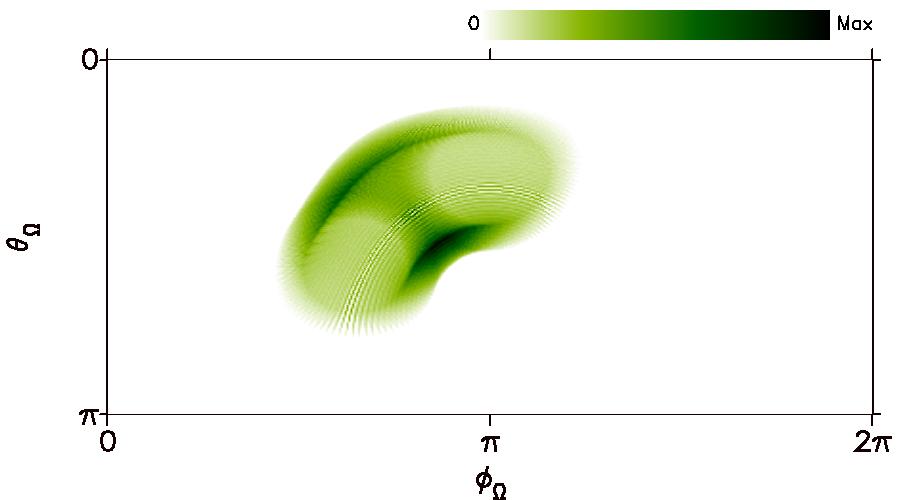}
	\includegraphics[width=.45\textwidth]{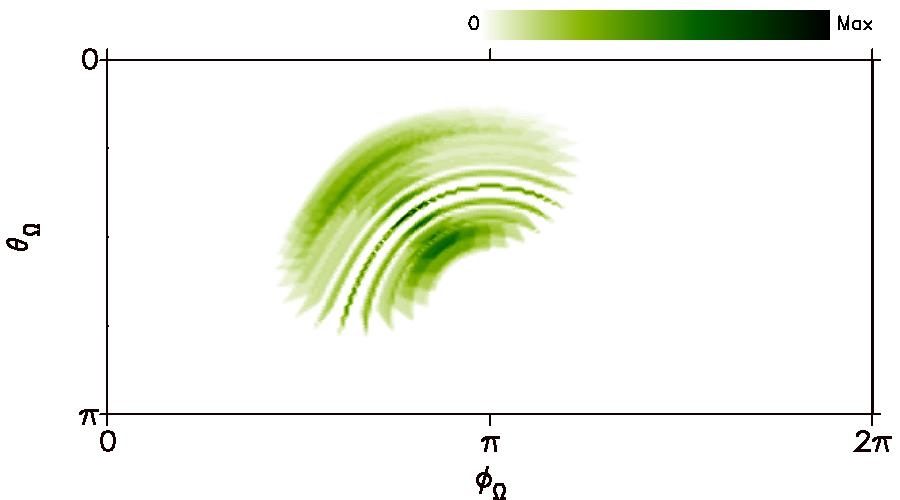}
	\includegraphics[width=.45\textwidth]{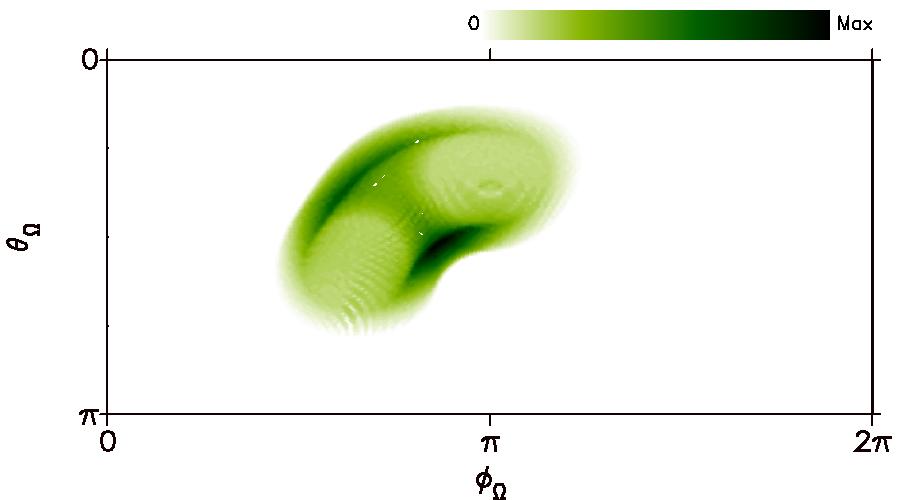}
	\includegraphics[width=.45\textwidth]{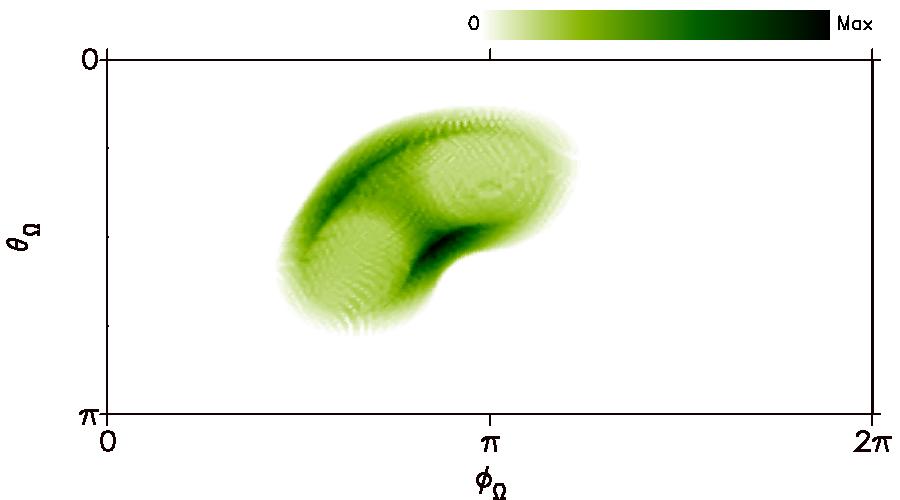}
	\caption{Effects of imbalanced sampling for the model OP1, with variations of only one resolution compared to the baseline combination $\bar N_p = \bar N_\lambda=\bar N_\chi=\bar N_l=100$. From top to bottom: $\bar N = 200$, $\bar N_\chi =25$,  $\bar N_\lambda =25$,  $\bar N_l =25$. On the other hand, increasing the ratio of any $\bar N_i/\bar N_p$ does not bring any visual effect.}
	\label{fig:test_geo_convergence2}
\end{figure}

As a numerical convergence test, we consider different resolutions, fixing $\bar N_p=\bar N_\lambda=\bar N_\chi=\bar N_l$ and considering the model OP1 of Table~\ref{tab:models}. In Fig.~\ref{fig:test_geo_convergence} we show the effect of increasing all the sampling numbers simultaneously. In Fig.~\ref{fig:test_geo_convergence2} we change one by one the sampling numbers, compared to a baseline one. Sampling effects include the appearance of intereference-like patterns, due to the low values of one or more physical sampling numbers $\bar N_\lambda,\bar N_\chi,\bar N_l$ compared to the number of patches in the sky, $2 \bar N_p^2$.

We stress that the geometry plays an important role, so that each model has to be tuned, depending on which sampling numbers are more important. Generally speaking, if a sampling number is associated to a spreading of radiation over the space (for instance, due to a large value of $\alpha$ which require a good sampling $\bar N_\chi$, or due to a wide range of values of $z_t$ along the trajectory, which requires a good sampling $\bar N_x$). A safe rule-of-thumb is to take $\bar N_\lambda=\bar N_\chi=\bar N_l = 4 \bar N_p \ge 200$: with this recipe, the variations of the physical parameters are fairly sampled and the sky maps do not present noise, converging to a smooth image. Note that cusps and sharp feature in the map can still be present and be resolved.

\label{lastpage}
\end{document}